\pdfoutput=1
\documentclass{article}

\usepackage{arxiv}

\usepackage[utf8]{inputenc} 
\usepackage[T1]{fontenc}    
\usepackage{lmodern}        
\usepackage{hyperref}       
\usepackage{url}            
\usepackage{booktabs}       
\usepackage{amsfonts}       
\usepackage{nicefrac}       
\usepackage{microtype}      
\usepackage{graphicx}

\title{\pkg{ergm} 4: New Features}

\author{
    Pavel N. Krivitsky
   \\
    School of Mathematics and Statistics \\
    University of New South Wales \\
  Sydney, NSW, Australia \\
  \texttt{\href{mailto:p.krivitsky@unsw.edu.au}{\nolinkurl{p.krivitsky@unsw.edu.au}}} \\
   \And
    David R. Hunter
   \\
    Department of Statistics \\
    Penn State University \\
  State College, PA, USA \\
  \texttt{\href{mailto:dhunter@stat.psu.edu}{\nolinkurl{dhunter@stat.psu.edu}}} \\
   \And
    Martina Morris
   \\
    Departments of Sociology and Statistics \\
    University of Washington \\
  Seattle, WA, USA \\
  \texttt{\href{mailto:morrism@uw.edu}{\nolinkurl{morrism@uw.edu}}} \\
   \And
    Chad Klumb
   \\
    Departments of Sociology and Statistics \\
    University of Washington \\
  Seattle, WA, USA \\
  \texttt{\href{mailto:cklumb@uw.edu}{\nolinkurl{cklumb@uw.edu}}} \\
  }

\usepackage{color}
\usepackage{fancyvrb}

\DefineVerbatimEnvironment{Highlighting}{Verbatim}{commandchars=\\\{\}}
\usepackage{framed}
\definecolor{shadecolor}{RGB}{248,248,248}
\newenvironment{Shaded}{\begin{snugshade}}{\end{snugshade}}

\newcommand{\CommentTok}[1]{\textcolor[rgb]{0.56,0.35,0.01}{\textit{#1}}}

\newcommand{\ControlFlowTok}[1]{\textcolor[rgb]{0.13,0.29,0.53}{\textbf{#1}}}
\newcommand{\DataTypeTok}[1]{\textcolor[rgb]{0.13,0.29,0.53}{#1}}
\newcommand{\DecValTok}[1]{\textcolor[rgb]{0.00,0.00,0.81}{#1}}

\newcommand{\ErrorTok}[1]{\textcolor[rgb]{0.64,0.00,0.00}{\textbf{#1}}}

\newcommand{\FloatTok}[1]{\textcolor[rgb]{0.00,0.00,0.81}{#1}}

\newcommand{\KeywordTok}[1]{\textcolor[rgb]{0.13,0.29,0.53}{\textbf{#1}}}
\newcommand{\NormalTok}[1]{#1}
\newcommand{\OperatorTok}[1]{\textcolor[rgb]{0.81,0.36,0.00}{\textbf{#1}}}
\newcommand{\OtherTok}[1]{\textcolor[rgb]{0.56,0.35,0.01}{#1}}

\newcommand{\StringTok}[1]{\textcolor[rgb]{0.31,0.60,0.02}{#1}}

\providecommand{\tightlist}{%
  \setlength{\itemsep}{0pt}\setlength{\parskip}{0pt}}

\newlength{\cslhangindent}
\newlength{\csllabelwidth}
\newlength{\cslentryspacingunit} 
\newenvironment{cslreferences}%
  {\setlength{\parindent}{0pt}%
  \everypar{\setlength{\hangindent}{\cslhangindent}}\ignorespaces}%
  {\par}
 {
  \setlength{\parindent}{0pt}
  \ifodd #1
  \let\oldpar\par
  \def\par{\hangindent=\cslhangindent\oldpar}
  \fi
  \setlength{\parskip}{#2\cslentryspacingunit}
 }%
 {}
\usepackage{calc}

\usepackage{ifthen}
\usepackage{amsmath,amssymb,tikz,enumitem,bm,booktabs,etoolbox,array}
\setlist{noitemsep}
\usepackage[font=normalsize]{subfig}

\AtBeginEnvironment{Highlighting}{\footnotesize}
\makeatletter
\renewcommand{\verbatim@font}{\ttfamily\footnotesize}
\makeatother
\def\y{\mathbf{y}}
\def\Y{\mathbf{Y}}

\def\nactors{ n }
\def\actors{ N }
\def\cnmapel{\eta}
\def\cnmap{\bm{\cnmapel}}
\def\dcnmap{\cnmap'}
\def\nnatpar{ p }
\def\curvparel{\theta}
\def\curvpar{{\bm{\curvparel}}}
\newcommand{\curvpark}[1][k]{\curvparel_{#1}}

\def\natcurvpars{\bm{\Theta}_{\text{N}}}
\def\mle{\hat{\curvpar}}

\def\ncurvpar{ q }
\def\genstatel{g}
\def\genstats{\mathbf{\genstatel}}
\newcommand{\genstat}[1]{\genstatel_{\text{#1}}}

\def\dyadvals{\mathbb{S}}

\def\guessind{ t }

\def\setsub{\backslash}

\def\grad{\bm{\nabla}}

\DeclareMathOperator{\E}{E}

\DeclareMathOperator{\Geometric}{Geometric}

\DeclareMathOperator{\ERGM}{ERGM}

\DeclareMathOperator{\0}{\mathbf{0}}

\def\dysY{\mathbb{Y}}
\def\netsY{\mathcal{Y}}
\def\iid{{\stackrel{\mathrm{i.i.d.}}{\sim}}}

\DeclareMathOperator{\Poisson}{Poisson}

\DeclareMathOperator{\Prob}{Pr}

\def\llik{\ell}

\def\h{h}

\def\normc{\kappa}

\def\cy{\normc_{\netsY}}

\def\Ey{\E_{\netsY}}

\def\cheg{\normc_{\h,\cnmap,\genstats}}

\def\Ptyheg{\Prob_{\curvpar,\netsY,\h,\cnmap,\genstats}}

\def\Dyheg{\ERGM_{\netsY,\h,\cnmap,\genstats}}

\newcommand{\cY}[1]{\normc_{#1}}
\newcommand{\DY}[1]{\ERGM_{#1}}

\newcommand{\EY}[1]{\E_{#1}}

\def\ij{{i,j}}

\def\pij{{(i,j)}}

\def\ijdysY{{\pij\in\dysY}}
\def\ynetsY{{\y\in\netsY}}
\def\ypnetsY{{\y'\in\netsY}}
\def\sij{_{i,j}}

\def\sji{_{j,i}}

\def\Yij{Y\!\sij}
\def\yij{y\sij}

\def\yji{y\sji}

\def\Yy{\Y=\y}
\def\sobs{^{\text{obs}}}

\def\yobs{\y\sobs}

\def\guess{\curvpar^\guessind}

\newcommand{\natpar}[1][]{\cnmap#1(\curvpar)}

\newcommand{\dnatmle}[1][]{\dcnmap#1(\mle)}

\newcommand{\myexp}[1]{\exp\left(#1\right)}

\providecommand{\abs}[1]{\lvert#1\rvert}

\def\t{^\top}

\def\defeq{\stackrel{\text{def}}{=}}

\newcommand{\innerprod}[2]{#1\t#2}

\def\beq{\begin{equation}}
\def\eeq{\end{equation}}

\newcommand{\en}[3]{#1 #3 #2}
\newcommand{\enbc}[1]{\{ #1 \}}

\providecommand{\E}{\operatorname{E}}

\newcommand{\indf}[1]{\mathbb{I}\en\{\}{#1}}

\newcommand{\figref}[1]{Figure~\ref{#1}}
\newcommand{\secref}[1]{Section~\ref{#1}}

\def\binary{^{\text{B}}}

\providecommand\pkg[1]{\textbf{#1}}
\providecommand\proglang[1]{\textsf{#1}}

\def\statnet{\pkg{statnet}}

\begin{document}
\maketitle

\begin{abstract}
The \pkg{ergm} package supports the statistical analysis and simulation
of network data. It anchors the \statnet{} suite of packages for network
analysis in \proglang{R} introduced in a special issue in \emph{Journal
of Statistical Software} in 2008. This article provides an overview of
the new functionality in the 2021 release of \pkg{ergm} version 4. These
include more flexible handling of nodal covariates, term operators that
extend and simplify model specification, new models for networks with
valued edges, improved handling of constraints on the sample space of
networks, and estimation with missing edge data. We also identify the
new packages in the \statnet{} suite that extend \pkg{ergm}'s
functionality to other network data types and structural features and
the robust set of online resources that support the \statnet{}
development process and applications.
\end{abstract}


\hypertarget{introduction}{%
\section{Introduction}\label{introduction}}

\label{sec:Introduction}

The \statnet{} suite of packages for \proglang{R} (\proglang{R} Core
Team, 2021) was first introduced in 2008, in volume 24 of \emph{Journal
of Statistical Software}, a special issue devoted to \statnet{}.
Together, these packages, which had already gone through the maturing
process of multiple releases, provided an integrated framework for the
statistical analysis of network data: from data storage and
manipulation, to visualization, estimation and simulation. Since that
time the existing packages have undergone continual updates to improve
and add capabilities, and many new packages have been added to extend
the range of network data that can be modeled (e.g., dynamic, valued,
sampled, multilevel). It is the \pkg{ergm} package, however, that
provides the statistical foundation for all of the other modeling
packages in the \statnet{} suite. Version 4 of \pkg{ergm}, released in
2021, is a major upgrade, representing more than a decade of changes and
improvements since (Hunter, Handcock, et al., 2008). This paper
describes the many updates to the functionality of the package of
interest to end users. It is a companion to (Krivitsky et al., 2022),
which discusses computational improvements in \pkg{ergm} version 4.

The exponential-family random graph model (ERGM) is a general
statistical framework for modeling the probability of a link (or tie)
between nodes in a network. It is implemented by the \pkg{ergm} package
and most of its related packages in the \statnet{} suite. We consider
networks over a set of nodes \(\actors=\{1,2,\dotsc,\nactors\}\). If
\(\dysY\subseteq \actors\times\actors\) denotes a set of potential
pairwise relationships among them, a binary network sample space can be
regarded as \(\netsY\subseteq 2^\dysY\), a subset of the power set of
potential relationships. More generally, we can define \(\dyadvals\) to
be a (possibly multivariate) set of possible relationship values. Then,
the sample space \(\netsY\subseteq \dyadvals^\dysY\) is a set whose
elements are of the form \(\{\Yij : \ijdysY \}\), where each \(\Yij\),
which we will call a dyad, maps the node pair \(\ijdysY\) into
\(\dyadvals\) and denotes the value of the relationship of \(\ijdysY\).

We begin by briefly presenting the fully general ERGM framework,
referring interested readers to Schweinberger et al. (2020) for
additional technical details. A random network \(\Y\) is distributed
according to an ERGM, written \(\Y\sim\Dyheg(\curvpar)\), if
\begin{equation}\label{ergm}
\Ptyheg(\Yy) = \frac{\h(\y)\exp\en\{\}{\natpar\t\genstats(\y)}} {\cheg(\curvpar,\netsY)},\ \ynetsY.
\end{equation} In Equation \eqref{ergm}, \(\netsY\) is the sample space
of networks; \(\curvpar\) is a \(\ncurvpar\)-dimensional parameter
vector; \(\h(\y)\) is a reference measure, typically a constant in the
case of binary ERGMs; \(\cnmap\) is a mapping from \(\curvpar\) to the
\(\nnatpar\)-vector of canonical parameters, a linear mapping in
non-curved ERGMs, and commonly an identity mapping
\(\natpar\equiv\curvpar\); \(\genstats\) is a \(\nnatpar\)-vector of
sufficient statistics; and \(\cheg(\curvpar,\netsY)\) is the normalizer
given by
\(\sum_{\ypnetsY} \h(\y') \exp\en\{\} {\natpar\t\genstats(\y')}\), which
is often intractable for models that seek to reproduce the dependence
across ties induced by social effects such as triadic closure. The
\emph{natural parameter space} of the model is
\(\natcurvpars \defeq \{\curvpar: \cheg(\curvpar,\netsY) < \infty\}\).

For a particular network application, one would typically employ a
special case of the fully general Equation \eqref{ergm}. For instance,
for binary ERGMs we typically define \(h(\y)\) to be a constant, as
discussed in \secref{sec:ReferenceSpecification}, in which case the
sufficient statistics can be thought of as modifying a uniform baseline
distribution over the potentially observable networks. Many of the
features of \pkg{ergm} and the related packages that comprise the
\statnet{} suite address the statistical complications that arise from
modeling network data using special cases of the ERGM in Equation
\eqref{ergm}.

In particular, the statistical framework implemented in \pkg{ergm} is
computationally intensive for models that specify dyadic dependence,
when \(\Ptyheg(\Yy)\) cannot be decomposed into a product of simple
functions of \(\yij\). In this case, the package relies on a central
Markov chain Monte Carlo (MCMC) algorithm for estimation and simulation,
along with maximum pseudo-likelihood estimation, contrastive divergence,
and simulated annealing in some contexts. Substantial improvements have
been made to all of these algorithms, producing efficiency and speed
gains of up to several orders of magnitude (Krivitsky et al., 2022).
This article describes the most important new capabilities that have
been added to \pkg{ergm} and its related packages since volume 24 of
\emph{Journal of Statistical Software} appeared in 2008. This includes
both the capabilities introduced in the version 4 release itself and in
releases 2.2.0--3.10.4, which postdate the \emph{JoSS} volume. Versions
in which each new capability was introduced can be obtained by running
\texttt{news(package="ergm")}.

In the examples throughout the article, we assume the reader is familiar
with the basic syntax and features of \pkg{ergm} included in the 2008
\emph{JoSS} volume. In some cases we demonstrate new, more general,
functionality by comparison, using the old syntax and the new to produce
the same result, then moving on with the new syntax to demonstrate the
additional utilities.

The source code for the latest version of the \pkg{ergm} package, along
with the \texttt{LICENSE} information under GPL-3, is available at
\url{https://github.com/statnet/ergm}.

\hypertarget{extension-packages-in-the-statnet-suite}{%
\section{Extension packages in the statnet
suite}\label{extension-packages-in-the-statnet-suite}}

\label{sec:Extensions}

The statistical models supported by the \statnet{} suite have been
extended by a growing number of new packages that provide additional
functionality in the general ERGM framework. While the focus of this
article is the base \pkg{ergm} package, in this section we provide a
brief overview of the extension packages and their specific
applications. Open source package development is on GitHub under the
\href{https://github.com/statnet}{\statnet{} organization}. Online
tutorials, found at \url{https://github.com/statnet/Workshops/wiki},
exist for \pkg{ergm} and many of these extension packages, and most
packages also include extended vignettes. Some of the key extension
packages, and the resources that support them, include:

\begin{description}
\item[Building custom terms for models]
One of the unique aspects of this modeling framework is that each
network statistic in an ERGM requires a specialized algorithm for
computing the value of the statistic from the data. The \pkg{ergm}
package has over 150 of the most common terms encoded---see
\texttt{vignette(\textquotesingle{}ergm-term-crossRef\textquotesingle{})}
or \texttt{help("ergmTerm")} for the full list---but the existing terms
are a small subset of the possible terms one can use in an ERGM. For
those who need a custom term, the package \pkg{ergm.userterms} (Hunter
et al., 2013) is designed to simplify the process of coding up new terms
for use in ERG model specification. Online workshop materials provide an
overview of the process, and demonstrate the use of this package (Hunter
\& Goodreau, 2019).
\item[Modeling temporal (dynamic) network data]
The \statnet{} suite contains several packages that provide a robust
framework for storing, visualizing, describing and modeling temporal
network data: The \pkg{networkDynamic} package extends \pkg{network} to
provide data storage and management utilities, the \pkg{tsna} package
extends \pkg{sna} (Butts, 2008) to provide descriptive statistics for
network objects that change over time, the \pkg{ndtv} package provides a
wide range of utilities for visualizing dynamic networks and saving both
static and animated output in standard formats, and \pkg{tergm} extends
\pkg{ergm} to fit the class of separable temporal ERGMs, from both
sampled and fully observed network data (Krivitsky \& Handcock, 2014).
There are two online workshops that demonstrate these tools: one that
demonstrates a typical workflow from data inspection to temporal
modeling (Morris \& Krivitsky, 2015), and another that focuses on
descriptive analyses and visualization (Bender-deMoll, 2016).
\item[Modeling valued edges]
The \pkg{ergm} itself contains a framework for modeling real-valued
edges (see \secref{sec:Valued} and \secref{sec:Operator}). Several other
packages provide specialized components for specific types of valued
edges: \pkg{ergm.count} for counts, \pkg{ergm.rank} for ordered
categories. The relevant theory supporting these packages may be found
in Krivitsky (2012) and Krivitsky \& Butts (2017), respectively.
\pkg{latentnet} for latent space models also supports non-binary
responses, although in a somewhat different manner (Krivitsky et al.,
2009; Krivitsky \& Handcock, 2008). Package vignettes and online
workshop materials provide an overview of the theory, and demonstrate
the use of these packages (Krivitsky \& Butts, 2019).
\item[Working with egocentrically sampled network data]
In the social and health sciences, egocentrically sampled network data
is the most common form of data available, because it can be collected
using standard sample survey methods. The \pkg{ergm.ego} package
provides methods for estimating ERGMs from egocentrically sampled
network data, with a principled framework for statistical inference. The
theory and an application of these methods may be found in Krivitsky \&
Morris (2017). Online workshop materials provide an overview of the
framework and demonstrate the use of the package (Morris \& Krivitsky,
2019).
\item[Multimode, multilayer, and multilevel networks]
In the social sciences, it is increasingly common to collect and fit
ERGMs on data on multiple relationship types (Krivitsky et al., 2020;
Wang, 2012) and ensembles of networks (Slaughter \& Koehly, 2016). These
capabilities are implemented in an extension package \pkg{ergm.multi}.
We refer the reader to the package manual and workshops for further
information.
\item[Modeling diffusion and epidemics on networks]
One of the most active application areas for ERGMs and TERGMs is in the
field of epidemic modeling. The \pkg{EpiModel} package is built on the
\statnet{} platform, and provides a unique set of tools for
statistically principled modeling of epidemics on networks (Jenness et
al., 2018). A robust set of online training materials is available at
\href{http://www.epimodel.org}{the EpiModel website}.
\end{description}

\hypertarget{sec:node-attr}{%
\section{Enhanced handling of nodal covariates}\label{sec:node-attr}}

Version 4 of \pkg{ergm} standardizes and provides greater flexibility
for handling covariates used by terms in an ERGM. In particular, these
covariates can be modified ``on-the-fly'' during model specification. A
vignette called \texttt{nodal\_attributes} is included in the package
and illustrates some of the new capabilities.

Here, we describe some of these enhancements using \pkg{ergm}'s
\texttt{faux.mesa.high} dataset, a simulated in-school friendship
network based on data collected on 205 students. We will focus on the
\texttt{Grade} attribute, an ordinal categorical variable with values 7
through 12 that can be accessed via the \texttt{\%v\%} operator:

\begin{Shaded}
\begin{Highlighting}[]
\KeywordTok{data}\NormalTok{(faux.mesa.high)}
\NormalTok{(faux.mesa.high }\OperatorTok{\%v\%}\StringTok{ "Grade"}\NormalTok{)[}\DecValTok{1}\OperatorTok{:}\DecValTok{20}\NormalTok{]  }\CommentTok{\# Look at first 20 nodes\textquotesingle{} (students\textquotesingle{}) grade levels}
\end{Highlighting}
\end{Shaded}

\begin{verbatim}
##  [1]  7  7 11  8 10 10  8 11  9  9  9 11  9 11  8 10 10  7 10  7
\end{verbatim}

Grade level is typical of the kind of covariate used to model selective
mixing in social networks: different hypotheses lead to different model
specifications. \pkg{ergm} 4 provides greater flexibility than earlier
versions of \pkg{ergm} to easily define and explore different
specifications.

We will sometimes call \texttt{summary()} and other times call
\texttt{ergm()} to demonstrate the functionality and output below.

\hypertarget{transformations-of-covariates}{%
\subsection{Transformations of
covariates}\label{transformations-of-covariates}}

\label{sec:Transformations}

It is sometimes desirable to specify a transformation of a nodal
attribute as a covariate in a model term. Most \texttt{ergm} terms now
support a new user interface, inspired by \pkg{purrr} (Henry \& Wickham,
2020), to specify transformations on one or more nodal attributes. Terms
typically use this new interface via arguments called \texttt{attr},
\texttt{attrs}, \texttt{by}, or \texttt{on}; the interpretation of the
argument depends on its type:

\begin{description}
\item[character string]
Extract the vertex attribute with this name.
\item[character vector of length greater than 1]
Extract the vertex attributes and paste them together, separated by dots
if the term expects categorical attributes and (typically) combine into
a covariate matrix if it expects quantitative attributes.
\item[function]
The function is called on the network on the left-hand side of the main
\texttt{ergm} formula and is expected to return a vector or matrix of
appropriate dimension. (Shorter vectors and matrix columns will be
recycled as needed.)
\item[formula]
Borrowing the interface from \pkg{tidyverse}, the expression on the
right hand side of the formula is evaluated in an environment of the
vertex attributes of the network, expected to return a vector or matrix
of appropriate dimension. (Shorter vectors and matrix columns will be
recycled as needed.) Within this expression, the network itself is
accessible as either \texttt{.} or \texttt{.nw}.
\item[\texttt{AsIs} object created by \texttt{I()}]
Use as is, checking only for correct length and type, with optional
attribute \texttt{"name"} providing the predictor's name.
\end{description}

For instance, here are three ways to compute the value of
\[\genstatel(\y) = \sum_{\ijdysY} \yij(\text{Grade}_i + \text{Grade}_j),\]
which in an ERGM may be interpreted as the linear effect of grade on
overall activity of an actor:

\begin{Shaded}
\begin{Highlighting}[]
\KeywordTok{summary}\NormalTok{(faux.mesa.high }\OperatorTok{\textasciitilde{}}\StringTok{ }\KeywordTok{nodecov}\NormalTok{(}\StringTok{"Grade"}\NormalTok{)                    }\CommentTok{\# String}
                       \OperatorTok{+}\StringTok{ }\KeywordTok{nodecov}\NormalTok{(}\OperatorTok{\textasciitilde{}}\NormalTok{Grade)                     }\CommentTok{\# Formula}
                       \OperatorTok{+}\StringTok{ }\KeywordTok{nodecov}\NormalTok{(}\ControlFlowTok{function}\NormalTok{(nw) nw}\OperatorTok{\%v\%}\StringTok{"Grade"}\NormalTok{)) }\CommentTok{\# Function}
\end{Highlighting}
\end{Shaded}

\begin{verbatim}
##        nodecov.Grade        nodecov.Grade nodecov.nw%v%"Grade" 
##                 3491                 3491                 3491
\end{verbatim}

Here is a more complicated formula-based use of \texttt{nodecov}, where
the first statistic is \[\genstatel(\y) = \sum_{\ijdysY} \yij
\left(
\frac{\left|\text{Grade}_i - \overline{\text{Grade}} \right|}{n}+
\frac{\left|\text{Grade}_j - \overline{\text{Grade}}\right|}{n}
\right),\] and \(n\) is the number of nodes, i.e., the network size, of
the network:

\begin{Shaded}
\begin{Highlighting}[]
\KeywordTok{summary}\NormalTok{(faux.mesa.high }\OperatorTok{\textasciitilde{}}\StringTok{ }\KeywordTok{nodecov}\NormalTok{(}\OperatorTok{\textasciitilde{}}\KeywordTok{abs}\NormalTok{(Grade }\OperatorTok{{-}}\StringTok{ }\KeywordTok{mean}\NormalTok{(Grade)) }\OperatorTok{/}\StringTok{ }\KeywordTok{network.size}\NormalTok{(.))}
                       \OperatorTok{+}\StringTok{ }\KeywordTok{nodecov}\NormalTok{(}\OperatorTok{\textasciitilde{}}\NormalTok{(Grade }\OperatorTok{{-}}\StringTok{ }\KeywordTok{mean}\NormalTok{(Grade)) }\OperatorTok{/}\StringTok{ }\KeywordTok{network.size}\NormalTok{(.)))}
\end{Highlighting}
\end{Shaded}

\begin{verbatim}
## nodecov.abs(Grade-mean(Grade))/network.size(.)    nodecov.(Grade-mean(Grade))/network.size(.) 
##                                      2.8565140                                     -0.2637716
\end{verbatim}

The non-zero output of the second statistic above, which omits the
absolute value, may be counterintuitive if you are expecting it to
return the sample mean grade, \(\overline{\text{Grade}}\). Node factor
statistics, however, are not the sample mean grade: each node is not
counted exactly once, but rather the number of cases it contributes is
equal to its degree.

Taking advantage of \texttt{nodecov}'s new ability to take matrix-valued
arguments, we might also evaluate a polynomial effect of \texttt{Grade},
as in the following quadratic example:\footnote{For this and other
  summaries, we omit the call information, deviances, and significance
  stars in the interests of space. The full summary information can be
  obtained by omitting \texttt{coef()} around the \texttt{summary()}
  call.}

\begin{Shaded}
\begin{Highlighting}[]
\KeywordTok{coef}\NormalTok{(}\KeywordTok{summary}\NormalTok{(}\KeywordTok{ergm}\NormalTok{(faux.mesa.high }\OperatorTok{\textasciitilde{}}\StringTok{ }\NormalTok{edges }\OperatorTok{+}\StringTok{ }\KeywordTok{nodecov}\NormalTok{(}\OperatorTok{\textasciitilde{}}\KeywordTok{cbind}\NormalTok{(Grade, }\DataTypeTok{Grade2=}\NormalTok{Grade}\OperatorTok{\^{}}\DecValTok{2}\NormalTok{)))))}
\end{Highlighting}
\end{Shaded}

\begin{verbatim}
##                  Estimate Std. Error MCMC %   z value     Pr(>|z|)
## edges           8.7297963 3.52880543      0  2.473867 0.0133659343
## nodecov.Grade  -1.4597723 0.39614405      0 -3.684953 0.0002287445
## nodecov.Grade2  0.0768836 0.02154632      0  3.568294 0.0003593133
\end{verbatim}

In the code above, the column for \texttt{Grade\^{}2} is explicitly
named \texttt{Grade2} whereas the column for \texttt{Grade} is named
implicitly by \proglang{R} itself. Omitting the name for a column not
otherwise named by \proglang{R} would result in a warning, as it is good
practice to name all variables in the model.

Alternatively, we can use \texttt{stats::poly} for orthogonal
polynomials. Here, the test for significance of the quadratic term is
identical to the non-orthogonal example, up to rounding error (though
the estimate is different given the orthogonal specification):

\begin{Shaded}
\begin{Highlighting}[]
\KeywordTok{coef}\NormalTok{(}\KeywordTok{summary}\NormalTok{(}\KeywordTok{ergm}\NormalTok{(faux.mesa.high }\OperatorTok{\textasciitilde{}}\StringTok{ }\NormalTok{edges }\OperatorTok{+}\StringTok{ }\KeywordTok{nodecov}\NormalTok{(}\OperatorTok{\textasciitilde{}}\KeywordTok{poly}\NormalTok{(Grade, }\DecValTok{2}\NormalTok{)))))}
\end{Highlighting}
\end{Shaded}

\begin{verbatim}
##                          Estimate Std. Error MCMC %    z value     Pr(>|z|)
## edges                   -4.662459 0.07309281      0 -63.788207 0.0000000000
## nodecov.poly(Grade,2).1 -1.207241 0.68018706      0  -1.774866 0.0759199607
## nodecov.poly(Grade,2).2  2.512615 0.70416949      0   3.568196 0.0003594477
\end{verbatim}

We can even pass a nodal covariate that is not already contained in the
network object. This example randomly generates a binary-valued nodal
covariate and sets its \texttt{name} attribute to be used as a label:

\begin{Shaded}
\begin{Highlighting}[]
\KeywordTok{set.seed}\NormalTok{(}\DecValTok{123}\NormalTok{)  }\CommentTok{\# Make exact output reproducible}
\NormalTok{randomcov \textless{}{-}}\StringTok{ }\KeywordTok{structure}\NormalTok{(}\KeywordTok{rbinom}\NormalTok{(}\KeywordTok{network.size}\NormalTok{(faux.mesa.high), }\DecValTok{1}\NormalTok{, }\FloatTok{0.5}\NormalTok{), }\DataTypeTok{name =} \StringTok{"random"}\NormalTok{)}
\KeywordTok{summary}\NormalTok{(faux.mesa.high }\OperatorTok{\textasciitilde{}}\StringTok{ }\KeywordTok{nodefactor}\NormalTok{(}\KeywordTok{I}\NormalTok{(randomcov)))}
\end{Highlighting}
\end{Shaded}

\begin{verbatim}
## nodefactor.random.1 
##                 199
\end{verbatim}

This syntax therefore allows for simulation or estimation of models with
inputs taken from arbitrary R functions or data sources, facilitating
the incorporation of ERGMs into more general tool chains.

\hypertarget{coding-categorical-attributes}{%
\subsection{Coding categorical
attributes}\label{coding-categorical-attributes}}

\label{sec:LevelSelection}

For model terms that use categorical attributes, \pkg{ergm} 4 has
extended the methods for selecting and/or transforming levels via the
use of the argument \texttt{levels}. Some terms, such as the
\texttt{sender} and \texttt{receiver} statistics of the \(p_1\) model
(Holland \& Leinhardt, 1981) and the corresponding \texttt{sociality}
statistics for undirected networks, treat the node labels themselves as
a categorical attribute. These terms use the \texttt{nodes=} argument,
rather than the \texttt{levels=} argument, to select a subset of the
nodes.

Typically, \texttt{levels} or \texttt{nodes} has a default that is
sensible for the term in question. (Information about the defaults of a
term \texttt{{[}name{]}} may be obtained by typing
\texttt{help("{[}name{]}-ergmTerm")} or \texttt{ergmTerm?{[}name{]}}.)
Interpretation of the possible values of the \texttt{levels} and
\texttt{nodes} arguments is available by typing
\texttt{help(nodal\_attributes)}. This interpretation is summarized as
follows:

\begin{description}
\item[\texttt{AsIs} object created by \texttt{I()}]
Use the given level, list of levels, or vector of levels as is.
\item[numeric or logical vector]
Used for indexing of a list of all possible levels (typically, unique
values of the attribute) in default order (typically lexicographic).
Logical values are recycled to the length of the vector indexed. In
particular, \texttt{levels=TRUE} retains all levels. Negative values
exclude. To specify numeric or logical levels literally, wrap them in
\texttt{I()}.
\item[\texttt{NULL}]
Retain all possible levels; usually equivalent to passing \texttt{TRUE}.
\item[character vector]
Use the given level(s) as is.
\item[function]
The function is called in an environment in which the network itself is
accessible as \texttt{.nw}, the list of unique values of the attribute
as \texttt{.} or as \texttt{.levels}, and the attribute vector itself as
\texttt{.attr}. Its return value is interpreted as above.
\item[formula]
The expression on the right hand side of the formula is evaluated in an
environment in which the network itself is accessible as \texttt{.nw},
the list of unique values of the attribute as \texttt{.} or as
\texttt{.levels}, and the attribute vector itself as \texttt{.attr}. Its
return value is interpreted as above.
\item[predefined function]
For convenience, a number of useful functions have been predefined.
\texttt{LARGEST}, which refers to the most frequent category, so, say,
to set such a category as the baseline, pass \texttt{levels=-LARGEST}.
\texttt{LARGEST(n)} will refer to the \texttt{n} largest categories.
\texttt{SMALLEST} works analogously, and ties in frequencies are broken
arbitrarily.
\end{description}

Returning to the \texttt{faux.mesa.high} example, we may treat
\texttt{Grade} as a categorical variable even though its values are
numeric. We see that \texttt{Grade} has six levels, numbered from 7 to
12:

\begin{Shaded}
\begin{Highlighting}[]
\KeywordTok{table}\NormalTok{(faux.mesa.high }\OperatorTok{\%v\%}\StringTok{ "Grade"}\NormalTok{)}
\end{Highlighting}
\end{Shaded}

\begin{verbatim}
## 
##  7  8  9 10 11 12 
## 62 40 42 25 24 12
\end{verbatim}

We may exclude the three smallest levels or, equivalently, include
levels 7, 8, and 9. Below are five of the myriad ways to do this in the
context of computing basic categorical effects on node activity,
implemented by \texttt{nodefactor}. In the second expression,
\texttt{I()} is necessary so that \texttt{7:9} is not treated as a
vector of indices.

\begin{Shaded}
\begin{Highlighting}[]
\KeywordTok{summary}\NormalTok{(faux.mesa.high }\OperatorTok{\textasciitilde{}}\StringTok{ }\KeywordTok{nodefactor}\NormalTok{(}\OperatorTok{\textasciitilde{}}\NormalTok{Grade, }\DataTypeTok{levels =} \OperatorTok{{-}}\KeywordTok{SMALLEST}\NormalTok{(}\DecValTok{3}\NormalTok{)))}
\end{Highlighting}
\end{Shaded}

\begin{verbatim}
## nodefactor.Grade.7 nodefactor.Grade.8 nodefactor.Grade.9 
##                153                 75                 65
\end{verbatim}

\begin{Shaded}
\begin{Highlighting}[]
\KeywordTok{summary}\NormalTok{(faux.mesa.high }\OperatorTok{\textasciitilde{}}\StringTok{ }\KeywordTok{nodefactor}\NormalTok{(}\OperatorTok{\textasciitilde{}}\NormalTok{Grade, }\DataTypeTok{levels =} \KeywordTok{I}\NormalTok{(}\DecValTok{7}\OperatorTok{:}\DecValTok{9}\NormalTok{)))}
\end{Highlighting}
\end{Shaded}

\begin{verbatim}
## nodefactor.Grade.7 nodefactor.Grade.8 nodefactor.Grade.9 
##                153                 75                 65
\end{verbatim}

\begin{Shaded}
\begin{Highlighting}[]
\KeywordTok{summary}\NormalTok{(faux.mesa.high }\OperatorTok{\textasciitilde{}}\StringTok{ }\KeywordTok{nodefactor}\NormalTok{(}\OperatorTok{\textasciitilde{}}\NormalTok{Grade, }\DataTypeTok{levels =} \KeywordTok{c}\NormalTok{(}\StringTok{"7"}\NormalTok{, }\StringTok{"8"}\NormalTok{, }\StringTok{"9"}\NormalTok{)))}
\end{Highlighting}
\end{Shaded}

\begin{verbatim}
## nodefactor.Grade.7 nodefactor.Grade.8 nodefactor.Grade.9 
##                153                 75                 65
\end{verbatim}

\begin{Shaded}
\begin{Highlighting}[]
\KeywordTok{summary}\NormalTok{(faux.mesa.high }\OperatorTok{\textasciitilde{}}\StringTok{ }\KeywordTok{nodefactor}\NormalTok{(}\StringTok{"Grade"}\NormalTok{, }\DataTypeTok{levels =} \ControlFlowTok{function}\NormalTok{(a) a }\OperatorTok{\%in\%}\StringTok{ }\DecValTok{7}\OperatorTok{:}\DecValTok{9}\NormalTok{))}
\end{Highlighting}
\end{Shaded}

\begin{verbatim}
## nodefactor.Grade.7 nodefactor.Grade.8 nodefactor.Grade.9 
##                153                 75                 65
\end{verbatim}

\begin{Shaded}
\begin{Highlighting}[]
\KeywordTok{summary}\NormalTok{(faux.mesa.high }\OperatorTok{\textasciitilde{}}\StringTok{ }\KeywordTok{nodefactor}\NormalTok{(}\StringTok{"Grade"}\NormalTok{, }\DataTypeTok{levels =} \OperatorTok{\textasciitilde{}}\NormalTok{. }\OperatorTok{\%in\%}\StringTok{ }\DecValTok{7}\OperatorTok{:}\DecValTok{9}\NormalTok{))}
\end{Highlighting}
\end{Shaded}

\begin{verbatim}
## nodefactor.Grade.7 nodefactor.Grade.8 nodefactor.Grade.9 
##                153                 75                 65
\end{verbatim}

Any of the arguments of \secref{sec:Transformations} may also be wrapped
in \texttt{COLLAPSE\_SMALLEST(attr,\ n,\ into)}, a convenience function
that will transform the attribute by collapsing the \texttt{n} least
frequent categories into one, naming it according to the \texttt{into}
argument where \texttt{into} must be of the same type (numeric,
character, etc.) as the vertex attribute in question. Consider the
\texttt{Race} factor of the \texttt{faux.mesa.high} network, where we
use \texttt{levels=TRUE} to display all levels since the default is
\texttt{levels=-1}:

\begin{Shaded}
\begin{Highlighting}[]
\KeywordTok{summary}\NormalTok{(faux.mesa.high }\OperatorTok{\textasciitilde{}}\StringTok{ }\KeywordTok{nodefactor}\NormalTok{(}\StringTok{"Race"}\NormalTok{, }\DataTypeTok{levels =} \OtherTok{TRUE}\NormalTok{))}
\end{Highlighting}
\end{Shaded}

\begin{verbatim}
## nodefactor.Race.Black  nodefactor.Race.Hisp nodefactor.Race.NatAm nodefactor.Race.Other 
##                    26                   178                   156                     1 
## nodefactor.Race.White 
##                    45
\end{verbatim}

Because the \texttt{Hisp} and \texttt{NatAm} categories are so much
larger than the other three categories in this network, we may wish to
combine the \texttt{Black}, \texttt{White}, and \texttt{Other}
categories. The code below accomplishes this using
\texttt{COLLAPSE\_SMALLEST} while also demonstrating how to use the
\pkg{magrittr} package's pipe function, \texttt{\%\textgreater{}\%}, for
improved readability:

\begin{Shaded}
\begin{Highlighting}[]
\KeywordTok{library}\NormalTok{(magrittr)}
\KeywordTok{summary}\NormalTok{(faux.mesa.high }\OperatorTok{\textasciitilde{}}\StringTok{ }\KeywordTok{nodefactor}\NormalTok{((}\OperatorTok{\textasciitilde{}}\NormalTok{Race) }\OperatorTok{\%\textgreater{}\%}
\StringTok{    }\KeywordTok{COLLAPSE\_SMALLEST}\NormalTok{(}\DecValTok{3}\NormalTok{, }\StringTok{"BWO"}\NormalTok{), }\DataTypeTok{levels =} \OtherTok{TRUE}\NormalTok{))}
\end{Highlighting}
\end{Shaded}

\begin{verbatim}
##   nodefactor.Race.BWO  nodefactor.Race.Hisp nodefactor.Race.NatAm 
##                    72                   178                   156
\end{verbatim}

\hypertarget{mixing-matrices}{%
\subsection{Mixing matrices}\label{mixing-matrices}}

Mixing matrices, which refer to the cross-tabulation of all edges by the
categorical attributes of the two nodes, are a common feature in models
that seek to represent selective mixing. The \texttt{mm} model term,
which stands for ``mixing matrix,'' generalizes the familiar
\texttt{nodemix} term from the original \pkg{ergm} implementation for
this purpose. Like \texttt{nodemix}, \texttt{mm} creates statistics
consisting of the cells of a matrix of counts in which the columns and
rows correspond to the levels of two categorical nodal covariates. For
\texttt{mm}, however, these covariates may or may not be the same,
making it more general. We use it here to demonstrate the
\texttt{levels2} argument.

Typing \texttt{help("mm-ergmTerm")} or, equivalently,
\texttt{ergmTerm?mm}, shows that the binary-network version of the term
takes the form \texttt{mm(attrs,\ levels=NULL,\ levels2=-1)}. The
\texttt{attrs} argument is a two-sided formula where the left and right
sides are the rows and columns, respectively, of the mixing matrix; if
only a one-sided formula or attribute name is given then the rows and
columns are taken to be the same. The optional \texttt{levels} argument
can similarly be a one- or two-sided formula, and it specifies the
levels of the row and column variables to keep. Finally, the optional
\texttt{levels2} argument may be used to select only a subset of the
matrix of statistics resulting from \texttt{attrs} and \texttt{levels}.

Using this functionality, we may specify custom mixing patterns that
depend upon attribute values. For instance, if we believe that the break
between junior high school (grades 7--9) and high school (grades 10--12)
creates a barrier to friendships across the boundary, we can create an
indicator variable \texttt{Grade} \(\ge10\), then compute a mixing
matrix on that variable using \texttt{mm} using a single call:

\begin{Shaded}
\begin{Highlighting}[]
\CommentTok{\# Mixing between lower and upper grades, with default specification:}
\KeywordTok{summary}\NormalTok{(faux.mesa.high }\OperatorTok{\textasciitilde{}}\StringTok{ }\KeywordTok{mm}\NormalTok{(}\OperatorTok{\textasciitilde{}}\NormalTok{Grade }\OperatorTok{\textgreater{}=}\StringTok{ }\DecValTok{10}\NormalTok{))}
\end{Highlighting}
\end{Shaded}

\begin{verbatim}
## mm[Grade>=10=FALSE,Grade>=10=TRUE]  mm[Grade>=10=TRUE,Grade>=10=TRUE] 
##                                 27                                 43
\end{verbatim}

\begin{Shaded}
\begin{Highlighting}[]
\CommentTok{\# Mixing with levels2 modified:}
\KeywordTok{summary}\NormalTok{(faux.mesa.high }\OperatorTok{\textasciitilde{}}\StringTok{ }\KeywordTok{mm}\NormalTok{(}\OperatorTok{\textasciitilde{}}\NormalTok{Grade }\OperatorTok{\textgreater{}=}\StringTok{ }\DecValTok{10}\NormalTok{, }\DataTypeTok{levels2 =} \OtherTok{NULL}\NormalTok{))}
\end{Highlighting}
\end{Shaded}

\begin{verbatim}
## mm[Grade>=10=FALSE,Grade>=10=FALSE]  mm[Grade>=10=FALSE,Grade>=10=TRUE] 
##                                 133                                  27 
##   mm[Grade>=10=TRUE,Grade>=10=TRUE] 
##                                  43
\end{verbatim}

The \texttt{Grade\textgreater{}=10} indicator variable is False (for
junior high school) and True (for high school), and with the undirected
friendships, this produces three possible combinations of the grade
indicator---False/False, False/True, and True/True. For the default
specification, \texttt{levels\ =\ NULL} keeps all levels of the
\texttt{Grade\textgreater{}=10} indicator variable and
\texttt{levels2\ =\ -1} eliminates the first statistic (False/False) in
the set of 3. For the modified specification, the
\texttt{levels2\ =\ NULL} argument keeps all of the statistics.

We can also use the \texttt{mm} formula interface to filter out certain
statistics from the full set of potential comparisons. An example from
the \texttt{nodal\_attributes} vignette within the \pkg{ergm} package
using the unmodified \texttt{Grade} attribute defines \texttt{levels2}
as a one-sided formula whose right side is a function that returns TRUE
or FALSE, depending on whether both elements of \texttt{.levels} ---the
list of values taken by a pair of nodes---are in the set
\texttt{c(7,\ 8)}. The example therefore captures mixing statistics only
involving students in grades 7 or 8:

\begin{Shaded}
\begin{Highlighting}[]
\KeywordTok{summary}\NormalTok{(faux.mesa.high }\OperatorTok{\textasciitilde{}}\StringTok{ }\KeywordTok{mm}\NormalTok{(}\StringTok{"Grade"}\NormalTok{, }\DataTypeTok{levels2 =} \OperatorTok{\textasciitilde{}}\KeywordTok{sapply}\NormalTok{(.levels,}
                           \ControlFlowTok{function}\NormalTok{(pair) pair[[}\DecValTok{1}\NormalTok{]] }\OperatorTok{\%in\%}\StringTok{ }\KeywordTok{c}\NormalTok{(}\DecValTok{7}\NormalTok{,}\DecValTok{8}\NormalTok{) }\OperatorTok{\&\&}\StringTok{ }\NormalTok{pair[[}\DecValTok{2}\NormalTok{]] }\OperatorTok{\%in\%}\StringTok{ }\KeywordTok{c}\NormalTok{(}\DecValTok{7}\NormalTok{,}\DecValTok{8}\NormalTok{))))}
\end{Highlighting}
\end{Shaded}

\begin{verbatim}
## mm[Grade=7,Grade=7] mm[Grade=7,Grade=8] mm[Grade=8,Grade=8] 
##                  75                   0                  33
\end{verbatim}

Finally, we give an example using two covariates, allowing us to capture
the tendency of sets of individuals defined by values of \texttt{Grade}
to mix with sets of individuals defined by values of \texttt{Race}:

\begin{Shaded}
\begin{Highlighting}[]
\KeywordTok{summary}\NormalTok{(faux.mesa.high }\OperatorTok{\textasciitilde{}}\StringTok{ }\KeywordTok{mm}\NormalTok{(Grade}\OperatorTok{\textgreater{}=}\DecValTok{10} \OperatorTok{\textasciitilde{}}\StringTok{ }\NormalTok{Race,}
                             \DataTypeTok{levels =} \OtherTok{TRUE} \OperatorTok{\textasciitilde{}}\StringTok{ }\KeywordTok{c}\NormalTok{(}\StringTok{"Hisp"}\NormalTok{, }\StringTok{"NatAm"}\NormalTok{, }\StringTok{"White"}\NormalTok{)))}
\end{Highlighting}
\end{Shaded}

\begin{verbatim}
##   mm[Grade>=10=TRUE,Race=Hisp] mm[Grade>=10=FALSE,Race=NatAm]  mm[Grade>=10=TRUE,Race=NatAm] 
##                             43                            115                             41 
## mm[Grade>=10=FALSE,Race=White]  mm[Grade>=10=TRUE,Race=White] 
##                             30                             15
\end{verbatim}

With all values of \texttt{Grade\textgreater{}=10} (i.e., False and
True) and three values of \texttt{Race} allowed according to the
\texttt{levels} argument, the full mixing matrix here would include
\(2\times 3\) statistics, though the default \texttt{levels2=-1} omits
the first of these so there is no
\texttt{Grade\textgreater{}=10=FALSE,Race=Hisp} statistic. When
interpreting mixing matrix effects of this type, bear in mind that two
covariates need not partition the vertex set in the same ways. Here, for
instance, there can be students both above and below grade 10 with each
race/ethnicity.

The \texttt{nodemix} term can do many of the same things that
\texttt{mm} can do. For both terms, \texttt{levels2} can take a matrix
as input; in particular, for \texttt{nodemix} this argument can take
character matrices to map multiple cells to the same statistic. For
instance, in the \texttt{faux.mesa.high} dataset, if we want to group
all sex-homophilous (male-male or female-female) ties together in the
same statistic while keeping the heterophilous (male-female) ties
separate, we can pass to \texttt{levels2} a \(2\times2\) matrix with
matching non-blank entries along the diagonal and blanks off the
diagonal:

\begin{Shaded}
\begin{Highlighting}[]
\NormalTok{m \textless{}{-}}\StringTok{ }\KeywordTok{matrix}\NormalTok{(}\KeywordTok{c}\NormalTok{(}\StringTok{"homophilous"}\NormalTok{, }\StringTok{""}\NormalTok{, }\StringTok{""}\NormalTok{, }\StringTok{"homophilous"}\NormalTok{), }\DecValTok{2}\NormalTok{, }\DecValTok{2}\NormalTok{)}
\KeywordTok{summary}\NormalTok{(faux.mesa.high }\OperatorTok{\textasciitilde{}}\StringTok{ }\KeywordTok{nodemix}\NormalTok{(}\StringTok{"Sex"}\NormalTok{, }\DataTypeTok{levels2 =}\NormalTok{ m))}
\end{Highlighting}
\end{Shaded}

\begin{verbatim}
## mix.Sex.homophilous         mix.Sex.F.M 
##                 132                  71
\end{verbatim}

\hypertarget{term-operators}{%
\section{Term operators}\label{term-operators}}

\label{sec:Operator}

\pkg{ergm} 4 introduces a new way to augment an \texttt{ergm} function
call that we call a \emph{term operator}, or simply \emph{operator}. In
mathematics, an operator is a function, like differentiation, that takes
functions as its inputs; analogously, a term operator takes one or more
ERGM formulas as input and transforms them by modifying their inputs
and/or outputs. Most operators therefore have a general form
\texttt{X(formula,\ ...)} where \texttt{X} is the name of the operator,
typically capitalized, \texttt{formula} is a one-sided formula
specifying the network statistics to be evaluated, and the remaining
arguments control the transformation applied to the network before
\texttt{formula} is evaluated and/or to the transformation applied to
the network statistics obtained by evaluating \texttt{formula}.
Operators are documented alongside other terms, accessible as
\texttt{help("{[}name{]}-ergmTerm")} or \texttt{ergmTerm?{[}name{]}},
and we describe some frequently used operators below.

\hypertarget{network-filters}{%
\subsection{Network filters}\label{network-filters}}

\label{sec:NetworkFilters}

Several operators allow the user to evaluate model terms on filtered
versions of the network, i.e., on particular subsets of the existing
nodes and/or edges.

\hypertarget{filtering-edges}{%
\subsubsection{Filtering edges}\label{filtering-edges}}

The operator \texttt{F(formula,\ filter)} evaluates the terms in
\texttt{formula} on a filtered network, with filtering specified by
\texttt{filter}. Here, \texttt{filter} is the right-hand side of a
formula that must contain one binary dyad-independent \texttt{ergm}
term, having exactly one statistic with a dyadwise contribution of 0 for
a 0-valued dyad. That is, the term must be expressible as
\begin{equation}\label{eq:DyadIndependent}
\genstatel(\y) = \sum_{\ijdysY} f_{i,j}(\yij),
\end{equation} where for all possible \((i,j)\), \(f_{i,j}(0)=0\). One
may verify that condition \eqref{eq:DyadIndependent} implies that an
ERGM containing the single term \(\genstatel(\y)\) has the property that
the dyads \(\Yij\) are jointly independent, which is why such a term is
called ``dyad-independent''. Examples of such terms include
\texttt{nodemix}, \texttt{nodematch}, \texttt{nodefactor},
\texttt{nodecov}, and \texttt{edgecov}. Then, \texttt{formula} will be
evaluated on a network constructed by taking \(\y\) and keeping only
those edges for which \(f_{i,j}(\yij) \ne 0\). This predicate can be
modified slightly by very simple comparison or logical expressions in
the \texttt{filter} formula. In particular, placing \texttt{!} in front
of the term negates it (i.e., keep \((i,j)\) only if
\(f_{i,j}(\yij) = 0\)) and comparison operators (\texttt{==},
\texttt{\textless{}}, etc.) allow comparing \(f_{i,j}(\yij)\) to values
other than 0.

\begin{figure}

{\centering \includegraphics[width=1\linewidth]{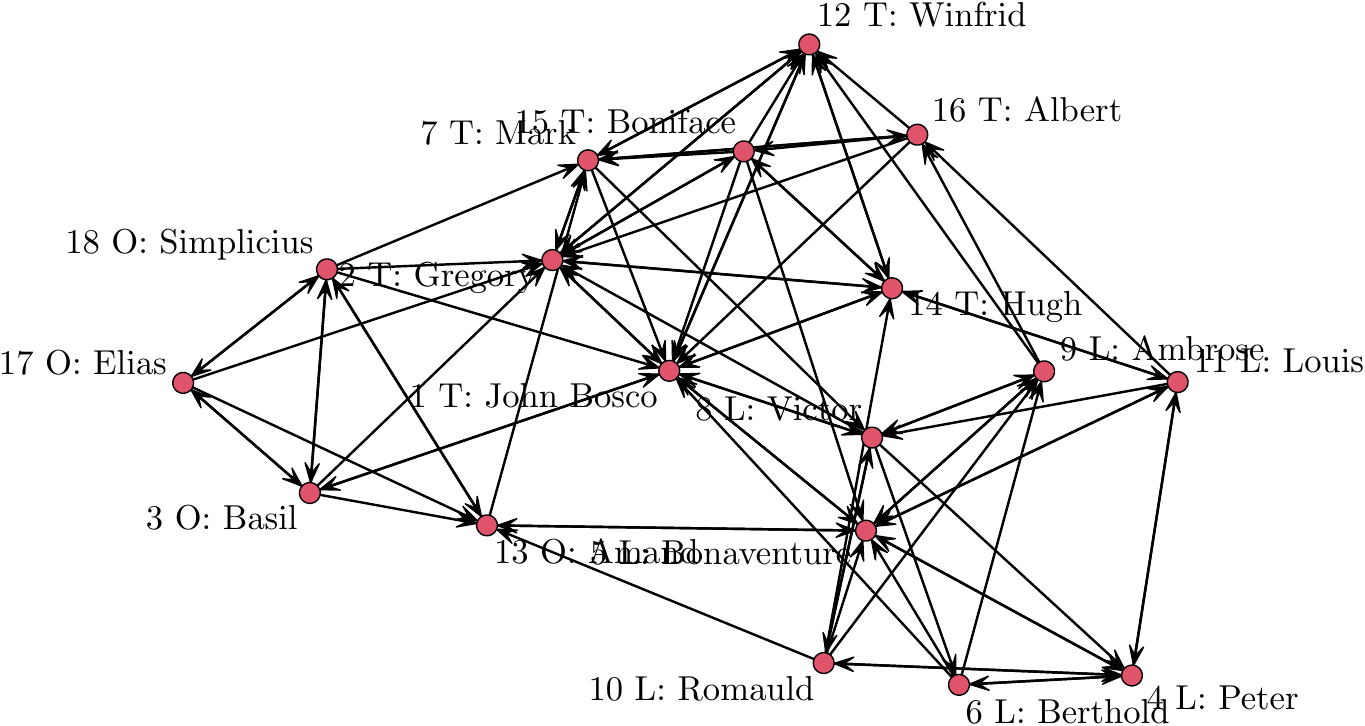} 

}

\caption{The monks dataset, with edges indicating directed liking relationships at any of three time points and nodes numbered from 1 to 18 and with group membership as assigned by Sampson indicated by L for Loyalists, O for Outcasts, and T for Young Turks.}\label{fig:sampson}
\end{figure}

Sampson's Monks (Sampson, 1968) can provide illustrative examples.
\pkg{ergm} includes a version of these data reporting cumulative liking
nominations over the three time periods Sampson asked a group of monks
to identify those they liked. This directed, 18-node network is depicted
in \figref{fig:sampson}.

\begin{Shaded}
\begin{Highlighting}[]
\KeywordTok{data}\NormalTok{(sampson)}
\NormalTok{lab \textless{}{-}}\StringTok{ }\KeywordTok{paste0}\NormalTok{(}\DecValTok{1}\OperatorTok{:}\DecValTok{18}\NormalTok{, }\StringTok{" "}\NormalTok{, }\KeywordTok{substr}\NormalTok{(samplike }\OperatorTok{\%v\%}\StringTok{ "group"}\NormalTok{, }\DecValTok{1}\NormalTok{, }\DecValTok{1}\NormalTok{), }\StringTok{": "}\NormalTok{, samplike }\OperatorTok{\%v\%}\StringTok{ "vertex.names"}\NormalTok{)}
\KeywordTok{plot}\NormalTok{(samplike, }\DataTypeTok{displaylabels =} \OtherTok{TRUE}\NormalTok{, }\DataTypeTok{label =}\NormalTok{ lab)}
\end{Highlighting}
\end{Shaded}

As an example of the \texttt{F} operator, the code below uses four
different methods to summarize the number of ties between pairs of nodes
in the \texttt{Turks} group in the \texttt{samplike} dataset:

\begin{Shaded}
\begin{Highlighting}[]
\KeywordTok{data}\NormalTok{(sampson)}
\KeywordTok{summary}\NormalTok{(samplike }\OperatorTok{\textasciitilde{}}\StringTok{ }\KeywordTok{nodematch}\NormalTok{(}\StringTok{"group"}\NormalTok{, }\DataTypeTok{diff=}\OtherTok{TRUE}\NormalTok{, }\DataTypeTok{levels=}\StringTok{"Turks"}\NormalTok{)}
                 \OperatorTok{+}\StringTok{ }\KeywordTok{F}\NormalTok{(}\OperatorTok{\textasciitilde{}}\KeywordTok{nodematch}\NormalTok{(}\StringTok{"group"}\NormalTok{), }\OperatorTok{\textasciitilde{}}\KeywordTok{nodefactor}\NormalTok{(}\StringTok{"group"}\NormalTok{, }\DataTypeTok{levels=}\StringTok{"Turks"}\NormalTok{))}
                 \OperatorTok{+}\StringTok{ }\KeywordTok{F}\NormalTok{(}\OperatorTok{\textasciitilde{}}\NormalTok{edges, }\OperatorTok{\textasciitilde{}}\KeywordTok{nodefactor}\NormalTok{(}\StringTok{"group"}\NormalTok{, }\DataTypeTok{levels=}\StringTok{"Turks"}\NormalTok{) }\OperatorTok{==}\StringTok{ }\DecValTok{2}\NormalTok{)}
                 \OperatorTok{+}\StringTok{ }\KeywordTok{F}\NormalTok{(}\OperatorTok{\textasciitilde{}}\NormalTok{edges, }\OperatorTok{\textasciitilde{}!}\KeywordTok{nodefactor}\NormalTok{(}\OperatorTok{\textasciitilde{}}\NormalTok{group}\OperatorTok{!=}\StringTok{"Turks"}\NormalTok{)))}
\end{Highlighting}
\end{Shaded}

\begin{verbatim}
##                                 nodematch.group.Turks 
##                                                    30 
## F(nodefactor("group",levels="Turks"))~nodematch.group 
##                                                    30 
##        F(nodefactor("group",levels="Turks")==2)~edges 
##                                                    30 
##                 F(!nodefactor(~group!="Turks"))~edges 
##                                                    30
\end{verbatim}

Here, the third method works because this particular \(f_{i,j}(\yij)\)
counts how many of the two nodes \(i\) and \(j\) are \texttt{Turks}, and
so equals 2 if and only if both are; and the fourth method works because
the new \(f_{i,j}(\yij)\) is 0 only if neither \(i\) nor \(j\) is a
non-\texttt{Turks} node.

It is also possible to filter on a quantitative variable. For instance,
an alternative way to count the number of edges in
\texttt{faux.mesa.high} that match on \texttt{"Grade"} is to report
total edges after filtering by node pairs whose absolute difference on
the \texttt{"Grade"} variable is less than 1:

\begin{Shaded}
\begin{Highlighting}[]
\KeywordTok{cbind}\NormalTok{(}\KeywordTok{summary}\NormalTok{(faux.mesa.high }\OperatorTok{\textasciitilde{}}\StringTok{ }\KeywordTok{nodematch}\NormalTok{(}\StringTok{"Grade"}\NormalTok{)),}
      \KeywordTok{summary}\NormalTok{(faux.mesa.high }\OperatorTok{\textasciitilde{}}\StringTok{ }\KeywordTok{F}\NormalTok{(}\OperatorTok{\textasciitilde{}}\NormalTok{edges, }\OperatorTok{\textasciitilde{}}\KeywordTok{absdiff}\NormalTok{(}\StringTok{"Grade"}\NormalTok{) }\OperatorTok{\textless{}}\StringTok{ }\DecValTok{1}\NormalTok{)))}
\end{Highlighting}
\end{Shaded}

\begin{verbatim}
##                 [,1] [,2]
## nodematch.Grade  163  163
\end{verbatim}

While \texttt{filter} must be dyad-independent, \texttt{formula} can
have dyad-dependent terms as well. For instance, we may count the
transitive triples---i.e., triples \((i,j,k)\) where
\(y_{i,j} = y_{j,k} = y_{i,k}=1\)---in the \texttt{samplike} network,
then perform the same count on the subnetwork consisting only of those
edges connecting two monks not in attendance in the minor seminary of
Cloisterville before coming to the monastery:

\begin{Shaded}
\begin{Highlighting}[]
\KeywordTok{summary}\NormalTok{(samplike }\OperatorTok{\textasciitilde{}}\StringTok{ }\NormalTok{ttriple }\OperatorTok{+}\StringTok{ }\KeywordTok{F}\NormalTok{(}\OperatorTok{\textasciitilde{}}\NormalTok{ttriple, }\OperatorTok{\textasciitilde{}}\KeywordTok{nodefactor}\NormalTok{(}\StringTok{"cloisterville"}\NormalTok{) }\OperatorTok{==}\StringTok{ }\DecValTok{0}\NormalTok{))}
\end{Highlighting}
\end{Shaded}

\begin{verbatim}
##                                   ttriple F(nodefactor("cloisterville")==0)~ttriple 
##                                       154                                        12
\end{verbatim}

\hypertarget{treating-directed-networks-as-undirected}{%
\subsubsection{Treating directed networks as
undirected}\label{treating-directed-networks-as-undirected}}

The operator \texttt{Symmetrize(formula,\ rule)} evaluates the terms in
\texttt{formula} on an undirected network constructed by symmetrizing
the underlying directed network according to \texttt{rule}. The possible
values of \texttt{rule}, which match the terminology of the
\texttt{symmetrize} function of the \pkg{sna} package, are (a) ``weak'',
(b) ``strong'', (c) ``upper'', and (d) ``lower''; for any \(i<j\), these
four values result in an undirected tie between \(i\) and \(j\) if and
only if (a) either \(\yij\) or \(\yji\) equals 1, (b) both \(\yij\) and
\(\yji\) equal 1, (c) \(\yij=1\), and (d) \(\yji=1\). For example,

\begin{Shaded}
\begin{Highlighting}[]
\KeywordTok{summary}\NormalTok{(samplike }\OperatorTok{\textasciitilde{}}\StringTok{ }\KeywordTok{Symmetrize}\NormalTok{(}\OperatorTok{\textasciitilde{}}\NormalTok{edges, }\StringTok{"strong"}\NormalTok{) }\OperatorTok{+}\StringTok{ }\NormalTok{mutual }\OperatorTok{+}\StringTok{ }\KeywordTok{Symmetrize}\NormalTok{(}\OperatorTok{\textasciitilde{}}\NormalTok{edges, }\StringTok{"weak"}\NormalTok{) }\OperatorTok{+}\StringTok{ }\NormalTok{edges)}
\end{Highlighting}
\end{Shaded}

\begin{verbatim}
## Symmetrize(strong)~edges                   mutual   Symmetrize(weak)~edges 
##                       28                       28                       60 
##                    edges 
##                       88
\end{verbatim}

will compute the number of node pairs \(i<j\) with reciprocated edges,
equivalent to mutuality, i.e., \(\yij=\yji=1\), along with the number of
node pairs in which at least one edge is present; summing these values
yields the total number of directed edges.

\hypertarget{extracting-subgraphs}{%
\subsubsection{Extracting subgraphs}\label{extracting-subgraphs}}

The operator \texttt{S(formula,\ attrs)} evaluates the terms in
\texttt{formula} on an induced subgraph constructed from vertices
identified by \texttt{attrs}. The \texttt{attrs} argument either takes a
value as explained in \secref{sec:LevelSelection} for the
\texttt{nodes=} argument or, to obtain a bipartite network, a two-sided
formula with the left-hand side specifying the tails and the right-hand
side specifying the heads. For instance, suppose that we wish to model
the density and mutuality dynamics within the group ``Young Turks'' as
different from those of the rest of the network:

\begin{Shaded}
\begin{Highlighting}[]
\KeywordTok{coef}\NormalTok{(}\KeywordTok{summary}\NormalTok{(}\KeywordTok{ergm}\NormalTok{(samplike}\OperatorTok{\textasciitilde{}}\NormalTok{edges }\OperatorTok{+}\StringTok{ }\NormalTok{mutual }\OperatorTok{+}\StringTok{ }\KeywordTok{S}\NormalTok{(}\OperatorTok{\textasciitilde{}}\NormalTok{edges }\OperatorTok{+}\StringTok{ }\NormalTok{mutual, }\OperatorTok{\textasciitilde{}}\NormalTok{(group }\OperatorTok{==}\StringTok{ "Turks"}\NormalTok{)),}
                  \DataTypeTok{control =} \KeywordTok{snctrl}\NormalTok{(}\DataTypeTok{seed =} \DecValTok{123}\NormalTok{))))}
\end{Highlighting}
\end{Shaded}

\begin{verbatim}
##                             Estimate Std. Error MCMC %   z value     Pr(>|z|)
## edges                      -2.007074  0.2377493      0 -8.441979 3.120095e-17
## mutual                      2.351613  0.4996500      0  4.706519 2.519819e-06
## S((group=="Turks"))~edges   2.812378  0.8650057      0  3.251282 1.148857e-03
## S((group=="Turks"))~mutual -2.165222  1.1965294      0 -1.809585 7.036009e-02
\end{verbatim}

Thus, the density within the group is statistically significantly
higher, whereas the reciprocation within the group is lower, though not
statistically significantly at the 5\%~level.

\begin{figure}

{\centering \includegraphics[width=0.7\textwidth]{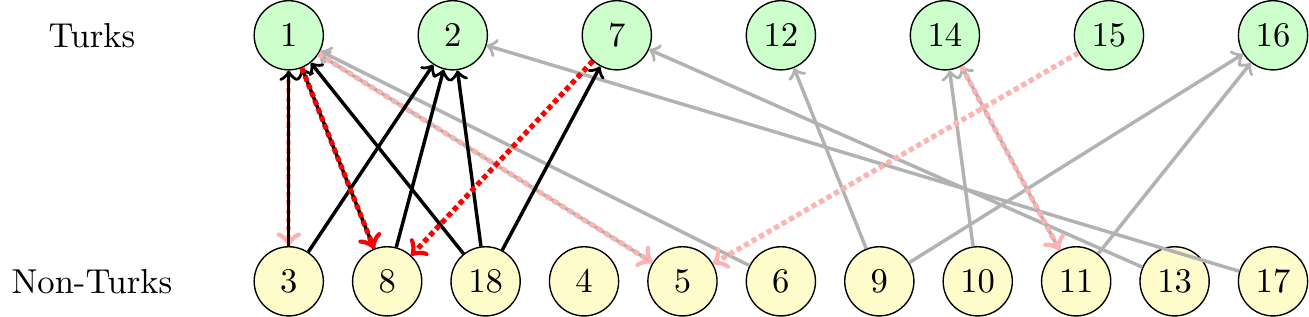} 

}

\caption{A bipartite induced subgraph between Turks (green) and Non-Turks (yellow). Edges involved in at least one undirected 4-cycle are emphasized. When directed edges from Non-Turks to Turks (black) are viewed as bipartite (undirected) edges, we obtain 4-cycles (3, 1, 18, 2), (3, 1, 8, 2), and (8, 1, 18, 2).  When directed edges from Turks to Non-Turks (dotted red) are also included, we obtain the additional 4-cycles (8, 1, 18, 7) and (8, 2, 18, 7).}\label{fig:FourCyclePlot}
\end{figure}

As another example, illustrated in \figref{fig:FourCyclePlot}, consider
the directed edges from non-Young Turks to Young Turks. Creating the
induced subgraph from these edges results in a bipartite network---which
is always taken to be undirected even though the edges were originally
directed---we may count the number of four-cycles:

\begin{Shaded}
\begin{Highlighting}[]
\KeywordTok{summary}\NormalTok{(samplike }\OperatorTok{\textasciitilde{}}\StringTok{ }\KeywordTok{S}\NormalTok{(}\OperatorTok{\textasciitilde{}}\KeywordTok{cycle}\NormalTok{(}\DecValTok{4}\NormalTok{), (group }\OperatorTok{!=}\StringTok{ "Turks"}\NormalTok{) }\OperatorTok{\textasciitilde{}}\StringTok{ }\NormalTok{(group }\OperatorTok{==}\StringTok{ "Turks"}\NormalTok{)))}
\end{Highlighting}
\end{Shaded}

\begin{verbatim}
## S((group!="Turks"),(group=="Turks"))~cycle4 
##                                           3
\end{verbatim}

On the other hand, if we treat the original network as undirected using
\texttt{Symmetrize} before creating the induced bipartite subgraph, we
see additional four-cycles. This example also illustrates that term
operators may be nested arbitrarily:

\begin{Shaded}
\begin{Highlighting}[]
\KeywordTok{summary}\NormalTok{(samplike }\OperatorTok{\textasciitilde{}}\StringTok{ }\KeywordTok{Symmetrize}\NormalTok{(}\OperatorTok{\textasciitilde{}}\KeywordTok{S}\NormalTok{(}\OperatorTok{\textasciitilde{}}\KeywordTok{cycle}\NormalTok{(}\DecValTok{4}\NormalTok{), (group }\OperatorTok{!=}\StringTok{ "Turks"}\NormalTok{) }\OperatorTok{\textasciitilde{}}\StringTok{ }\NormalTok{(group }\OperatorTok{==}\StringTok{ "Turks"}\NormalTok{)), }\StringTok{"weak"}\NormalTok{))}
\end{Highlighting}
\end{Shaded}

\begin{verbatim}
## Symmetrize(weak)~S((group!="Turks"),(group=="Turks"))~cycle4 
##                                                            5
\end{verbatim}

Finally, we illustrate a common use case in which \texttt{Symmetrize} is
used to analyze mutuality in a directed network as a function of a
predictor. The \texttt{faux.dixon.high} dataset is a directed friendship
network of seventh through twelfth graders. Suppose we wish to check how
strongly the tendency toward mutuality in friendships is affected by
students' closeness in grade level.

\begin{Shaded}
\begin{Highlighting}[]
\KeywordTok{data}\NormalTok{(faux.dixon.high)}
\NormalTok{FDHfit \textless{}{-}}\StringTok{ }\KeywordTok{ergm}\NormalTok{(faux.dixon.high }\OperatorTok{\textasciitilde{}}\StringTok{ }\NormalTok{edges }\OperatorTok{+}\StringTok{ }\NormalTok{mutual }\OperatorTok{+}\StringTok{ }\KeywordTok{absdiff}\NormalTok{(}\StringTok{"grade"}\NormalTok{)}
               \OperatorTok{+}\StringTok{ }\KeywordTok{Symmetrize}\NormalTok{(}\OperatorTok{\textasciitilde{}}\KeywordTok{absdiff}\NormalTok{(}\StringTok{"grade"}\NormalTok{), }\StringTok{"strong"}\NormalTok{), }\DataTypeTok{control=}\KeywordTok{snctrl}\NormalTok{(}\DataTypeTok{seed=}\DecValTok{321}\NormalTok{))}
\KeywordTok{coef}\NormalTok{(}\KeywordTok{summary}\NormalTok{(FDHfit))}
\end{Highlighting}
\end{Shaded}

\begin{verbatim}
##                                    Estimate Std. Error MCMC %    z value      Pr(>|z|)
## edges                            -3.2492290 0.04728446      0 -68.716639  0.000000e+00
## mutual                            3.2447618 0.11742536      0  27.632549 4.523720e-168
## absdiff.grade                    -0.9139858 0.04289319      0 -21.308413 9.485629e-101
## Symmetrize(strong)~absdiff.grade -0.4130642 0.17653795      0  -2.339804  1.929386e-02
\end{verbatim}

After correcting for the overall network density, the propensity for
friendships to be reciprocated, and the predictive effect of grade
difference on friendship formation, the difference in grade level has a
statistically significant negative effect on the tendency to form mutual
friendships (\(p\)-value = 0.019).

\hypertarget{interaction-effects}{%
\subsection{Interaction effects}\label{interaction-effects}}

\label{sec:InteractionEffects}

For binary ERGMs, interactions between dyad-independent \texttt{ergm}
terms can be specified in a manner similar to \texttt{lm} and
\texttt{glm} via the \texttt{:} and \texttt{*} operators. (See
\secref{sec:NetworkFilters} for a definition of dyad-independent.)

Let us first consider the colon (\texttt{:}) operator. Generally, if
term \texttt{A} creates \(p_{\texttt{A}}\) statistics and term
\texttt{B} creates \(p_{\texttt{B}}\) statistics, then \texttt{A:B} will
create \(p_{\texttt{A}} \times p_{\texttt{B}}\) new statistics. If
\texttt{A} and \texttt{B} are dyad-independent terms, expressed for
\(a=1,\dotsc,p_{\texttt{A}}\) and \(b=1,\dotsc, p_{\texttt{B}}\) as \[
g_{\texttt{A}}(\y)=\sum_{\ijdysY} x^{\texttt{A}}_{i,j}\yij
\text{\ and\ }
g_{\texttt{B}}(\y)=\sum_{\ijdysY} x^{\texttt{B}}_{i,j}\yij
\] for appropriate covariate matrices \(X^{\texttt{A}}\) and
\(X^{\texttt{B}}\), then the corresponding interaction term is
\begin{equation}\label{InteractionTerm}
g_{\texttt{A:B}}(\y)=\sum_{\ijdysY} x^{\texttt{A}}_{i,j}x^{\texttt{B}}_{i,j}\yij.
\end{equation}

As an example, consider the \texttt{Grade} and \texttt{Sex} effects,
expressed as model terms via \texttt{nodefactor}, in the
\texttt{faux.mesa.high} dataset:

\begin{Shaded}
\begin{Highlighting}[]
\KeywordTok{summary}\NormalTok{(faux.mesa.high }\OperatorTok{\textasciitilde{}}\StringTok{ }\KeywordTok{nodefactor}\NormalTok{(}\StringTok{"Grade"}\NormalTok{, }\DataTypeTok{levels =} \OtherTok{TRUE}\NormalTok{)}\OperatorTok{:}\KeywordTok{nodefactor}\NormalTok{(}\StringTok{"Sex"}\NormalTok{))}
\end{Highlighting}
\end{Shaded}

\begin{verbatim}
##  nodefactor.Grade.7:nodefactor.Sex.M  nodefactor.Grade.8:nodefactor.Sex.M 
##                                   70                                   99 
##  nodefactor.Grade.9:nodefactor.Sex.M nodefactor.Grade.10:nodefactor.Sex.M 
##                                   63                                   46 
## nodefactor.Grade.11:nodefactor.Sex.M nodefactor.Grade.12:nodefactor.Sex.M 
##                                   38                                   26
\end{verbatim}

In the call above, we deliberately include all \texttt{Grade}-factor
levels via \texttt{levels=TRUE}, whereas we employ the default behavior
of \texttt{nodefactor} for the \texttt{Sex} factor, which leaves out one
level. Thus, the 6-level \texttt{Grade} factor and the 2-level
\texttt{Sex} factor, with one level of the latter omitted, produce
\(6\times 1\) interaction terms in this example.

The \texttt{*} operator, by contrast, produces all interactions in
addition to the main effects or statistics. Therefore, in the scenario
described above, \texttt{A*B} will add
\(p_{\texttt{A}} + p_{\texttt{B}} + p_{\texttt{A}} \times p_{\texttt{B}}\)
statistics to the model. Below, we use the default behavior of
\texttt{nodefactor} on both the 6-level \texttt{Grade} factor and the
2-level \texttt{Sex} factor, together with an additional \texttt{edges}
term, to produce a model with \(1+ 5 + 1 + 5\times 1\) terms:

\begin{Shaded}
\begin{Highlighting}[]
\NormalTok{m \textless{}{-}}\StringTok{ }\KeywordTok{ergm}\NormalTok{(faux.mesa.high }\OperatorTok{\textasciitilde{}}\StringTok{ }\NormalTok{edges }\OperatorTok{+}\StringTok{ }\KeywordTok{nodefactor}\NormalTok{(}\StringTok{"Grade"}\NormalTok{) }\OperatorTok{*}\StringTok{ }\KeywordTok{nodefactor}\NormalTok{(}\StringTok{"Sex"}\NormalTok{))}
\KeywordTok{print}\NormalTok{(}\KeywordTok{summary}\NormalTok{(m), }\DataTypeTok{digits =} \DecValTok{3}\NormalTok{)}
\end{Highlighting}
\end{Shaded}

\begin{verbatim}
## Call:
## ergm(formula = faux.mesa.high ~ edges + nodefactor("Grade") * 
##     nodefactor("Sex"))
## 
## Maximum Likelihood Results:
## 
##                                      Estimate Std. Error MCMC % z value Pr(>|z|)    
## edges                                  -3.028      0.173      0  -17.53  < 1e-04 ***
## nodefactor.Grade.8                     -1.424      0.263      0   -5.41  < 1e-04 ***
## nodefactor.Grade.9                     -1.166      0.229      0   -5.10  < 1e-04 ***
## nodefactor.Grade.10                    -1.633      0.357      0   -4.58  < 1e-04 ***
## nodefactor.Grade.11                    -0.328      0.237      0   -1.38  0.16714    
## nodefactor.Grade.12                    -0.794      0.324      0   -2.45  0.01429 *  
## nodefactor.Sex.M                       -1.764      0.240      0   -7.36  < 1e-04 ***
## nodefactor.Grade.8:nodefactor.Sex.M     1.386      0.202      0    6.86  < 1e-04 ***
## nodefactor.Grade.9:nodefactor.Sex.M     1.012      0.211      0    4.79  < 1e-04 ***
## nodefactor.Grade.10:nodefactor.Sex.M    1.347      0.264      0    5.11  < 1e-04 ***
## nodefactor.Grade.11:nodefactor.Sex.M    0.419      0.240      0    1.75  0.08074 .  
## nodefactor.Grade.12:nodefactor.Sex.M    1.059      0.290      0    3.65  0.00026 ***
## ---
## Signif. codes:  0 '***' 0.001 '**' 0.01 '*' 0.05 '.' 0.1 ' ' 1
## 
##      Null Deviance: 28987  on 20910  degrees of freedom
##  Residual Deviance:  2189  on 20898  degrees of freedom
##  
## AIC: 2213  BIC: 2308  (Smaller is better. MC Std. Err. = 0)
\end{verbatim}

Equation \eqref{InteractionTerm} implies that the change statistic
corresponding to dyad \(\pij\) is given by
\(x^{\texttt{A}}_{i,j}x^{\texttt{B}}_{i,j}\); that is, the change
statistic for the interaction is the product of the change statistics.
One may define interaction change statistics for arbitrary pairs of
terms similarly---that is, by taking the interaction change statistic as
the product of the corresponding change statistics---though in the case
of dyad-dependent terms it is unclear that a change statistic obtained
as the product of change statistics corresponds to any ERGM sufficient
statistic in the sense of Equation \eqref{ergm}. Therefore, attempting
to create interactions involving dyad-dependent terms will create an
error by default in \pkg{ergm}. If one wishes to create such
interactions anyway, the default behavior may be changed using the
\texttt{interact.dependent} term option as described in
\secref{sec:term-options}. Interactions involving curved ERGM terms are
not supported in \pkg{ergm} 4.

Since interaction terms are defined by multiplying change statistics
dyadwise and then (for dyad-independent terms) summing over all dyads,
interactions of terms are not the same as products of those terms. For
instance, given a nodal covariate \texttt{"a"}, the interaction of
\texttt{nodecov("a")} with itself is different than the effect of the
square of the covariate, as we observe in the case of the
\texttt{wealth} covariate of the (undirected) Florentine marriage
dataset:

\begin{Shaded}
\begin{Highlighting}[]
\KeywordTok{data}\NormalTok{(florentine)}
\KeywordTok{summary}\NormalTok{(flomarriage }\OperatorTok{\textasciitilde{}}\StringTok{ }\KeywordTok{nodecov}\NormalTok{(}\StringTok{"wealth"}\NormalTok{)}\OperatorTok{:}\KeywordTok{nodecov}\NormalTok{(}\StringTok{"wealth"}\NormalTok{) }\OperatorTok{+}\StringTok{ }\KeywordTok{nodecov}\NormalTok{(}\OperatorTok{\textasciitilde{}}\NormalTok{wealth}\OperatorTok{\^{}}\DecValTok{2}\NormalTok{))}
\end{Highlighting}
\end{Shaded}

\begin{verbatim}
## nodecov.wealth:nodecov.wealth              nodecov.wealth^2 
##                        284058                        187814
\end{verbatim}

\hypertarget{reparametrizing-the-model}{%
\subsection{Reparametrizing the model}\label{reparametrizing-the-model}}

The term operator \texttt{Sum(formulas,\ label)} allows arbitrary linear
combinations of existing statistics to be added to the model. Suppose
\(\genstats_1(\y),\dotsc,\genstats_K(\y)\) is a set of \(K\)
vector-valued network statistics, each corresponding to one or more
\texttt{ergm} terms and of arbitrary dimension. Also suppose that
\(A_1,\dotsc,A_K\) is a set of known constant matrices all having the
same number of rows such that each matrix multiplication
\(A_k\genstats_k(\y)\) is well-defined. Then it is now possible to
define the statistic
\[\genstats_\text{Sum}(\y)=\sum_{k=1}^K A_k\genstats_k(\y).\]

The first argument to \texttt{Sum} is a formula or a list of \(K\)
formulas, each representing a vector statistic. If a formula has a
left-hand side, the left-hand side will be used to define the
corresponding \(A_k\) matrix: If it is a scalar or a vector, \(A_k\)
will be a diagonal matrix thus multiplying each element by its
corresponding element; and if it is a matrix, \(A_k\) will be used
directly. When no left-hand side is given, \(A_k\) is defined as 1. To
simplify this function for some common cases, if the left-hand side is
\texttt{"sum"} or \texttt{"mean"}, the sum (or mean) of the statistics
in the formula is calculated.

As an example, consider a vector of statistics consisting of the numbers
of friendship ties received by each subgroup of Sampson's monks:

\begin{Shaded}
\begin{Highlighting}[]
\KeywordTok{summary}\NormalTok{(samplike }\OperatorTok{\textasciitilde{}}\StringTok{ }\KeywordTok{nodeifactor}\NormalTok{(}\StringTok{"group"}\NormalTok{, }\DataTypeTok{levels =} \OtherTok{TRUE}\NormalTok{))}
\end{Highlighting}
\end{Shaded}

\begin{verbatim}
##    nodeifactor.group.Loyal nodeifactor.group.Outcasts    nodeifactor.group.Turks 
##                         29                         13                         46
\end{verbatim}

We may create a single statistic equal to the friendship ties received
by both groups of non-Outcasts by adding the first and third components
of the \texttt{nodefactor} vector, either by left-multiplying by
\(\begin{bmatrix} 1 & 0 & 1 \end{bmatrix}\) or by deselecting the second
component at the \texttt{nodeifactor} level and summing the remaining
two:

\begin{Shaded}
\begin{Highlighting}[]
\KeywordTok{summary}\NormalTok{(samplike }\OperatorTok{\textasciitilde{}}\StringTok{ }\KeywordTok{Sum}\NormalTok{(}\KeywordTok{cbind}\NormalTok{(}\DecValTok{1}\NormalTok{, }\DecValTok{0}\NormalTok{, }\DecValTok{1}\NormalTok{) }\OperatorTok{\textasciitilde{}}\StringTok{ }\KeywordTok{nodeifactor}\NormalTok{(}\StringTok{"group"}\NormalTok{, }\DataTypeTok{levels =} \OtherTok{TRUE}\NormalTok{), }\StringTok{"nf.L\_T"}\NormalTok{) }\OperatorTok{+}
\StringTok{                    }\KeywordTok{Sum}\NormalTok{(}\StringTok{"sum"} \OperatorTok{\textasciitilde{}}\StringTok{ }\KeywordTok{nodeifactor}\NormalTok{(}\StringTok{"group"}\NormalTok{, }\DataTypeTok{levels =} \DecValTok{{-}2}\NormalTok{), }\StringTok{"nf.L\_T"}\NormalTok{))}
\end{Highlighting}
\end{Shaded}

\begin{verbatim}
## Sum~nf.L_T Sum~nf.L_T 
##         75         75
\end{verbatim}

Whereas the \texttt{Sum} operator operates on network statistics,
\texttt{Parametrize(formula,\ params,\ map,\ gradient=NULL,\ minpar=-Inf,\ maxpar=+Inf,\ cov=NULL)}
operates on the parameters. The \texttt{formula} argument specifies a
vector statistic \(\genstats_k(\y)\) involving one or more terms and, if
curved terms are specified, a mapping \(\cnmap_k(\curvpar)\). The
remaining arguments follow the curved ERGM template: The function
\texttt{map} must take arguments \texttt{x}, \texttt{n}, and
\texttt{...} and map the parameter vector (whose length and names are
specified by the \texttt{params} argument) into the domain of
\(\cnmap_k\), transforming an ERGM term
\(\innerprod{\cnmap_k(\curvpar_k)}{\genstats_k(\y)}\) to
\(\innerprod{\cnmap_k(\cnmap_\star(\curvpar_k))}{\genstats_k(\y)}\),
where \(\cnmap_\star\) is the function specified by \texttt{map}. The
function \texttt{gradient} must take the same arguments as \texttt{map}
and return the gradient matrix, \texttt{minpar} and \texttt{maxpar}
specify the box constraints of the domain of \texttt{map}, and
\texttt{cov} provides an optional argument to \texttt{map}. If
\texttt{formula} is not curved, \(\cnmap_k(\curvpar)\) is simply the
identity function.

To simplify this function for some common special cases, if
\texttt{map="rep"}, the parameter vector will simply be replicated to
make it as long as required by \(\cnmap_k(\curvpar)\), and the gradient
will be evaluated automatically. Similarly, if the user is certain that
\texttt{map} is linear or affine, the gradient will be calculated
automatically if \texttt{gradient="linear"} is specified.

To illustrate this, consider a simple model with the baseline edge
effect and a single attractiveness effect for monks who are not
Outcasts. Following are four different ways to specify this model:

\begin{Shaded}
\begin{Highlighting}[]
\CommentTok{\# Calculated nodal covariate:}
\NormalTok{f1 \textless{}{-}}\StringTok{ }\NormalTok{samplike }\OperatorTok{\textasciitilde{}}\StringTok{ }\NormalTok{edges }\OperatorTok{+}\StringTok{ }\KeywordTok{nodeifactor}\NormalTok{(}\OperatorTok{\textasciitilde{}}\NormalTok{group }\OperatorTok{!=}\StringTok{ "Outcasts"}\NormalTok{)}
\KeywordTok{summary}\NormalTok{(f1)}
\end{Highlighting}
\end{Shaded}

\begin{verbatim}
##                              edges nodeifactor.group!="Outcasts".TRUE 
##                                 88                                 75
\end{verbatim}

\begin{Shaded}
\begin{Highlighting}[]
\CommentTok{\# Transform the statistic:}
\NormalTok{f2 \textless{}{-}}\StringTok{ }\NormalTok{samplike }\OperatorTok{\textasciitilde{}}\StringTok{ }\NormalTok{edges }\OperatorTok{+}
\StringTok{                 }\KeywordTok{Sum}\NormalTok{(}\KeywordTok{cbind}\NormalTok{(}\DecValTok{1}\NormalTok{,}\DecValTok{0}\NormalTok{,}\DecValTok{1}\NormalTok{) }\OperatorTok{\textasciitilde{}}\StringTok{ }\KeywordTok{nodeifactor}\NormalTok{(}\StringTok{"group"}\NormalTok{,}\DataTypeTok{levels=}\OtherTok{TRUE}\NormalTok{), }\StringTok{"nf.L\_T"}\NormalTok{)}
\KeywordTok{summary}\NormalTok{(f2)}
\end{Highlighting}
\end{Shaded}

\begin{verbatim}
##      edges Sum~nf.L_T 
##         88         75
\end{verbatim}

\begin{Shaded}
\begin{Highlighting}[]
\CommentTok{\# Transform the parameters:}
\NormalTok{f3 \textless{}{-}}\StringTok{ }\NormalTok{samplike }\OperatorTok{\textasciitilde{}}\StringTok{ }\NormalTok{edges }\OperatorTok{+}\StringTok{ }\KeywordTok{Parametrize}\NormalTok{(}\OperatorTok{\textasciitilde{}}\KeywordTok{nodeifactor}\NormalTok{(}\StringTok{"group"}\NormalTok{, }\DataTypeTok{levels=}\OtherTok{TRUE}\NormalTok{), }\StringTok{"nf.L\_T"}\NormalTok{,}
                               \ControlFlowTok{function}\NormalTok{(x,n,...) }\KeywordTok{c}\NormalTok{(x,}\DecValTok{0}\NormalTok{,x), }\DataTypeTok{gradient=}\StringTok{"linear"}\NormalTok{)}
\KeywordTok{summary}\NormalTok{(f3)}
\end{Highlighting}
\end{Shaded}

\begin{verbatim}
##                      edges    nodeifactor.group.Loyal nodeifactor.group.Outcasts 
##                         88                         29                         13 
##    nodeifactor.group.Turks 
##                         46
\end{verbatim}

\begin{Shaded}
\begin{Highlighting}[]
\CommentTok{\# Select groups, replicate parameter:}
\NormalTok{f4 \textless{}{-}}\StringTok{ }\NormalTok{samplike }\OperatorTok{\textasciitilde{}}\StringTok{ }\NormalTok{edges }\OperatorTok{+}\StringTok{ }\KeywordTok{Parametrize}\NormalTok{(}\OperatorTok{\textasciitilde{}}\KeywordTok{nodeifactor}\NormalTok{(}\StringTok{"group"}\NormalTok{, }\DataTypeTok{levels=}\OperatorTok{{-}}\DecValTok{2}\NormalTok{), }\StringTok{"nf.L\_T"}\NormalTok{, }\StringTok{"rep"}\NormalTok{)}
\KeywordTok{summary}\NormalTok{(f4)}
\end{Highlighting}
\end{Shaded}

\begin{verbatim}
##                   edges nodeifactor.group.Loyal nodeifactor.group.Turks 
##                      88                      29                      46
\end{verbatim}

We may now verify that all four fitted models return the same parameter
estimates:

\begin{Shaded}
\begin{Highlighting}[]
\KeywordTok{cbind}\NormalTok{(}\KeywordTok{coef}\NormalTok{(}\KeywordTok{ergm}\NormalTok{(f1)), }\KeywordTok{coef}\NormalTok{(}\KeywordTok{ergm}\NormalTok{(f2)), }\KeywordTok{coef}\NormalTok{(}\KeywordTok{ergm}\NormalTok{(f3)), }\KeywordTok{coef}\NormalTok{(}\KeywordTok{ergm}\NormalTok{(f4)))}
\end{Highlighting}
\end{Shaded}

\begin{verbatim}
##                                          [,1]       [,2]       [,3]       [,4]
## edges                              -1.4423838 -1.4423838 -1.4423838 -1.4423838
## nodeifactor.group!="Outcasts".TRUE  0.6661217  0.6661217  0.6661217  0.6661217
\end{verbatim}

Whereas the \texttt{Sum} operator calculates linear combinations of
network statistics, the \texttt{Prod} operator calculates the products
of their powers. As of this writing, it is implemented for positive
statistics only, by first applying the \texttt{Log} operator (which
returns the natural logarithm, \texttt{log} in \proglang{R}, of the
statistics passed to it), then the \texttt{Sum} operator, and finally
the \texttt{Exp} operator (which returns the exponential function,
\texttt{exp} in \proglang{R}). As a simple illustration, we may verify
that the \texttt{Sum} and \texttt{Prod} operators do in fact produce
network statistics as expected if we simply use each with a list of
formulas having no left hand side:

\begin{Shaded}
\begin{Highlighting}[]
\KeywordTok{summary}\NormalTok{(faux.dixon.high }\OperatorTok{\textasciitilde{}}\StringTok{ }\NormalTok{edges }\OperatorTok{+}\StringTok{ }\NormalTok{mutual }\OperatorTok{+}\StringTok{ }\KeywordTok{Sum}\NormalTok{(}\KeywordTok{list}\NormalTok{(}\OperatorTok{\textasciitilde{}}\NormalTok{edges, }\OperatorTok{\textasciitilde{}}\NormalTok{mutual), }\StringTok{"EdgesAndMutual"}\NormalTok{)}
        \OperatorTok{+}\StringTok{ }\KeywordTok{Prod}\NormalTok{(}\KeywordTok{list}\NormalTok{(}\OperatorTok{\textasciitilde{}}\NormalTok{edges, }\OperatorTok{\textasciitilde{}}\NormalTok{mutual), }\StringTok{"EdgesAndMutual"}\NormalTok{))}
\end{Highlighting}
\end{Shaded}

\begin{verbatim}
##                  edges                 mutual     Sum~EdgesAndMutual Exp~Sum~EdgesAndMutual 
##                   1197                    219                   1416                 262143
\end{verbatim}

\hypertarget{sample-space-constraints}{%
\section{Sample space constraints}\label{sample-space-constraints}}

\label{sec:constraints}

In \secref{sec:Introduction}, we saw that the sample space \(\netsY\) is
a subset of the power set \(2^\dysY\), where \(\dysY\) in itself a
subset of all potential relationships. Many applications in fact take
\(\dysY\) to be the set of all relationships and \(\netsY= 2^\dysY\),
but it is sometimes desirable to restrict the sample space by placing
constraints on which relationships \((i,j)\) are allowed in \(\dysY\)
and further which networks \(\y \in 2^\dysY\) are allowed in \(\netsY\).
As a simple example, a bipartite network allows only edges connecting
nodes from one subset, or mode, to nodes from its complement. This
particular constraint is so commonly used that it is hard-coded into
\pkg{network} and \pkg{ergm}. As another example, consider the inverse
of a bipartite setting, in which edges are only allowed \emph{within}
subsets of the node set, a situation often called a block-diagonal
constraint. As still another, some applications impose a cap on the
degree of any node, which constrains the sample space to include only
those networks in which every node has a permitted degree.

In all of the cases above, correct statistical inference for ERGMs
depends on correctly incorporating constraints into the fitting process.
They are specified using the \texttt{constraints} argument, a one-sided
formula whose terms specify the constraints on the sample space. For
example, \texttt{constraints\ =\ \textasciitilde{}edges} specifies
\(\netsY^{\text{\texttt{edges}}}=\enbc{\y'\in \netsY : \abs{\y'}=\abs{\y}}\),
where \(\y\) is the observed network, specified on the left-hand side.
Some constraints, such as \texttt{fixedas(y1,y0)} focus on constraining
\(\dysY\)---in this case, as
\(\dysY^{\text{\texttt{fixedas(y1,y0)}}}=\enbc{\ijdysY:\pij\in\y^1\land\pij\notin\y^0}\)---with
\(\netsY\equiv 2^\dysY\).

Multiple constraints can be specified on a formula, separated by
\texttt{+} to impose a new constraint in additional to prior or (in some
instances) \texttt{-} to relax preceding constraints. Earlier versions
of the \pkg{ergm} package implemented a number of constraints, as
described for example in Section 3 of Morris et al. (2008). Since that
time, many additional types of constraints and methods for imposing them
have been added, some of which we describe in this section. A full list
of currently implemented constraints is obtained via
\texttt{?\ ergmConstraint}, and a specific constraint
\texttt{{[}name{]}} can be looked up with
\texttt{help("{[}name{]}-ergmConstraint")} or
\texttt{ergmConstraint?{[}name{]}}.

\hypertarget{arbitrary-combinations-of-dyad-independent-constraints}{%
\subsection{Arbitrary combinations of dyad-independent
constraints}\label{arbitrary-combinations-of-dyad-independent-constraints}}

In general, every combination of constraints requires a somewhat
different Metropolis--Hastings proposal algorithm for efficient
sampling, and so it may be impractical support every possible
combination of constraints. \emph{Dyad-independent} constraints, which
affect \(\netsY\) only through \(\dysY\), and do not induce stochastic
dependencies among the dyad states, are an exception. These include
constraining specific dyads (such as the above-mentioned
\texttt{observed} and \texttt{fixedas} constraints), dyads incident on
specific actors (such as the \texttt{egocentric} constraint), and
block-diagonal structure; and \emph{any} combination of dyad-independent
constraints is a dyad-independent constraint. For some such
combinations, \pkg{ergm} and other packages provide optimized
implementations. For the rest, \pkg{ergm} falls back to a general but
efficient implementation that uses run-length encoding tools provided by
package \pkg{rle} (Krivitsky, 2020) to efficiently store sets of
non-constrained dyads and rejection sampling to efficiently select a
dyad for the proposal.

Here, we illustrate some of \pkg{ergm}'s capabilities using a dataset
due to Coleman (1964) that is small enough that computational
inefficiency will not present problems. These data are self-reported
friendship ties among 73 boys measured at two time points during the
1957--1958 academic year and they are included as a
\(2\times73\times73\) array and documented in the \pkg{sna} package. We
use the Coleman data to create a \texttt{network} object with
\(2\times73\) nodes:

\begin{Shaded}
\begin{Highlighting}[]
\KeywordTok{library}\NormalTok{(sna)}
\KeywordTok{data}\NormalTok{(coleman)}
\NormalTok{cole \textless{}{-}}\StringTok{ }\KeywordTok{matrix}\NormalTok{(}\DecValTok{0}\NormalTok{, }\DecValTok{2} \OperatorTok{*}\StringTok{ }\DecValTok{73}\NormalTok{, }\DecValTok{2} \OperatorTok{*}\StringTok{ }\DecValTok{73}\NormalTok{)}
\CommentTok{\# Upper left and lower right blocks:}
\NormalTok{cole[}\DecValTok{1}\OperatorTok{:}\DecValTok{73}\NormalTok{, }\DecValTok{1}\OperatorTok{:}\DecValTok{73}\NormalTok{] \textless{}{-}}\StringTok{ }\NormalTok{coleman[}\DecValTok{1}\NormalTok{, , ]}
\NormalTok{cole[}\DecValTok{73} \OperatorTok{+}\StringTok{ }\NormalTok{(}\DecValTok{1}\OperatorTok{:}\DecValTok{73}\NormalTok{), }\DecValTok{73} \OperatorTok{+}\StringTok{ }\NormalTok{(}\DecValTok{1}\OperatorTok{:}\DecValTok{73}\NormalTok{)] \textless{}{-}}\StringTok{ }\NormalTok{coleman[}\DecValTok{2}\NormalTok{, , ]}
\CommentTok{\# Upper right and lower left blocks:}
\KeywordTok{diag}\NormalTok{(cole[}\DecValTok{1}\OperatorTok{:}\DecValTok{73}\NormalTok{, }\DecValTok{73} \OperatorTok{+}\StringTok{ }\NormalTok{(}\DecValTok{1}\OperatorTok{:}\DecValTok{73}\NormalTok{)]) \textless{}{-}}\StringTok{ }\KeywordTok{diag}\NormalTok{(cole[}\DecValTok{73} \OperatorTok{+}\StringTok{ }\NormalTok{(}\DecValTok{1}\OperatorTok{:}\DecValTok{73}\NormalTok{), }\DecValTok{1}\OperatorTok{:}\DecValTok{73}\NormalTok{]) \textless{}{-}}\StringTok{ }\DecValTok{1}
\NormalTok{ncole \textless{}{-}}\StringTok{ }\KeywordTok{network}\NormalTok{(cole)}
\NormalTok{ncole }\OperatorTok{\%v\%}\StringTok{ "Semester"}\NormalTok{ \textless{}{-}}\StringTok{ }\KeywordTok{rep}\NormalTok{(}\KeywordTok{c}\NormalTok{(}\StringTok{"Fall"}\NormalTok{, }\StringTok{"Spring"}\NormalTok{), }\DataTypeTok{each =} \DecValTok{73}\NormalTok{)}
\NormalTok{ncole}
\end{Highlighting}
\end{Shaded}

\begin{verbatim}
##  Network attributes:
##   vertices = 146 
##   directed = TRUE 
##   hyper = FALSE 
##   loops = FALSE 
##   multiple = FALSE 
##   bipartite = FALSE 
##   total edges= 652 
##     missing edges= 0 
##     non-missing edges= 652 
## 
##  Vertex attribute names: 
##     Semester vertex.names 
## 
## No edge attributes
\end{verbatim}

By construction, the \texttt{ncole} network includes the Fall 1957
semester data and the Spring 1958 data as the upper left \(73\times73\)
and lower right \(73\times73\) blocks, respectively. In addition, the
upper right and lower left blocks indicate which nodes are the same
person; that is, \(\yij=1\) whenever \(i\) and \(j\) are the same boy
measured at two different times. This latter information is redundant
because the ordering of the 73 boys is the same in both fall and spring,
yet we include it to illustrate some techniques using entries that are
not on the main block diagonal and because in principle it might not
always be the case that the same individuals are observed at both time
points.

\hypertarget{constraints-via-the-dyads-operator}{%
\subsection{Constraints via the Dyads
operator}\label{constraints-via-the-dyads-operator}}

\label{sec:Dyads}

The \texttt{Dyads(fix=NULL,\ vary=NULL)} operator takes one or two
\texttt{ergm} formulas that may contain only dyad-independent terms. For
the terms in the \texttt{fix=} formula, dyads that affect the network
statistic (i.e., have nonzero change statistic) for \emph{any} the terms
will be fixed at their current values. For the terms in the
\texttt{vary=} formula, only those that change \emph{at least one} of
the terms will be allowed to vary, and all others will be fixed. If both
formulas are given, the dyads that vary either for one or for the other
will be allowed to vary. A formula passed without an argument name will
default to \texttt{fix=}, for consistency with other constraints'
semantics.

The key to our treatment of the \texttt{ncole} network using the
\texttt{Dyads} operator is the \texttt{Semester} vertex attribute:

\begin{Shaded}
\begin{Highlighting}[]
\KeywordTok{table}\NormalTok{(ncole }\OperatorTok{\%v\%}\StringTok{ "Semester"}\NormalTok{)}
\end{Highlighting}
\end{Shaded}

\begin{verbatim}
## 
##   Fall Spring 
##     73     73
\end{verbatim}

In particular, the \texttt{nodematch("Semester")} term has a change
statistic equal to one for exactly those dyads representing boys
measured during the same semester, and this change statistic is zero
otherwise. Therefore, in our 146-node directed network there are
\(146\times 145\), or \(21{,}170\), total dyads, of which
\(2\times73\times72\), or \(10{,}512\), have nonzero change statistics
for \texttt{nodematch("Semester")}. We can easily see exactly how many
total edges there are and how many of these are in the upper left or
lower right blocks:

\begin{Shaded}
\begin{Highlighting}[]
\KeywordTok{summary}\NormalTok{(ncole }\OperatorTok{\textasciitilde{}}\StringTok{ }\NormalTok{edges }\OperatorTok{+}\StringTok{ }\KeywordTok{nodematch}\NormalTok{(}\StringTok{"Semester"}\NormalTok{))}
\end{Highlighting}
\end{Shaded}

\begin{verbatim}
##              edges nodematch.Semester 
##                652                506
\end{verbatim}

We can now calculate directly the log-odds, or logit, for both the block
diagonal and the off-block diagonal subnetworks, then verify that the
\texttt{Dyads} operator can accomplish these same calculations using a
constrained ERGM. First, we \texttt{fix} dyads with nonzero change
statistics, which corresponds to the block off-diagonal entries:

\begin{Shaded}
\begin{Highlighting}[]
\NormalTok{logit \textless{}{-}}\StringTok{ }\ControlFlowTok{function}\NormalTok{(p) }\KeywordTok{log}\NormalTok{(p}\OperatorTok{/}\NormalTok{(}\DecValTok{1}\OperatorTok{{-}}\NormalTok{p))}
\KeywordTok{cbind}\NormalTok{(}\KeywordTok{logit}\NormalTok{((}\DecValTok{652} \OperatorTok{{-}}\StringTok{ }\DecValTok{506}\NormalTok{) }\OperatorTok{/}\StringTok{ }\NormalTok{(}\DecValTok{21170} \OperatorTok{{-}}\StringTok{ }\DecValTok{10512}\NormalTok{)),}
      \KeywordTok{coef}\NormalTok{(}\KeywordTok{ergm}\NormalTok{(ncole }\OperatorTok{\textasciitilde{}}\StringTok{ }\NormalTok{edges, }\DataTypeTok{constraints =} \OperatorTok{\textasciitilde{}}\KeywordTok{Dyads}\NormalTok{(}\DataTypeTok{fix =} \OperatorTok{\textasciitilde{}}\KeywordTok{nodematch}\NormalTok{(}\StringTok{"Semester"}\NormalTok{)))))}
\end{Highlighting}
\end{Shaded}

\begin{verbatim}
##            [,1]      [,2]
## edges -4.276666 -4.276666
\end{verbatim}

Next, we allow dyads with nonzero change statistics to \texttt{vary},
which corresponds to the block diagonal entries:

\begin{Shaded}
\begin{Highlighting}[]
\KeywordTok{cbind}\NormalTok{(}\KeywordTok{logit}\NormalTok{(}\DecValTok{506} \OperatorTok{/}\StringTok{ }\DecValTok{10512}\NormalTok{),}
      \KeywordTok{coef}\NormalTok{(}\KeywordTok{ergm}\NormalTok{(ncole }\OperatorTok{\textasciitilde{}}\StringTok{ }\NormalTok{edges, }\DataTypeTok{constraints =} \OperatorTok{\textasciitilde{}}\KeywordTok{Dyads}\NormalTok{(}\DataTypeTok{vary =} \OperatorTok{\textasciitilde{}}\KeywordTok{nodematch}\NormalTok{(}\StringTok{"Semester"}\NormalTok{)))))}
\end{Highlighting}
\end{Shaded}

\begin{verbatim}
##            [,1]      [,2]
## edges -2.984404 -2.984404
\end{verbatim}

If we remove the constraints entirely, \texttt{ncole} has 652 edges out
of a possible \(21{,}170\):

\begin{Shaded}
\begin{Highlighting}[]
\KeywordTok{cbind}\NormalTok{(}\KeywordTok{logit}\NormalTok{(}\DecValTok{652}\OperatorTok{/}\DecValTok{21170}\NormalTok{), }\KeywordTok{coef}\NormalTok{(}\KeywordTok{ergm}\NormalTok{(ncole }\OperatorTok{\textasciitilde{}}\StringTok{ }\NormalTok{edges)))}
\end{Highlighting}
\end{Shaded}

\begin{verbatim}
##            [,1]      [,2]
## edges -3.449013 -3.449013
\end{verbatim}

A significant limitation of this specific constraint is that its
initialization requires testing every possible dyad and therefore takes
up time and memory in proportion to the square of the number of nodes.

\hypertarget{constraints-via-blocks}{%
\subsection{Constraints via blocks}\label{constraints-via-blocks}}

\label{sec:BlocksConstraint}

The \texttt{blocks} operator constrains changes to any dyads that
involve certain pairs of categories defined by a particular nodal
covariate. We may reproduce the examples of \secref{sec:Dyads} using
\texttt{blocks}. First, consider the full complement of statistics
produced by the \texttt{nodemix} model term:

\begin{Shaded}
\begin{Highlighting}[]
\KeywordTok{summary}\NormalTok{(ncole }\OperatorTok{\textasciitilde{}}\StringTok{ }\KeywordTok{nodemix}\NormalTok{(}\StringTok{"Semester"}\NormalTok{, }\DataTypeTok{levels =} \OtherTok{TRUE}\NormalTok{, }\DataTypeTok{levels2 =} \OtherTok{TRUE}\NormalTok{))}
\end{Highlighting}
\end{Shaded}

\begin{verbatim}
##     mix.Semester.Fall.Fall   mix.Semester.Spring.Fall   mix.Semester.Fall.Spring 
##                        243                         73                         73 
## mix.Semester.Spring.Spring 
##                        263
\end{verbatim}

The \texttt{levels\ =\ TRUE} argument ensures that nodemix considers
every value of \texttt{"group"} in constructing a mixing matrix of
possible dyad combinations. The \texttt{levels2\ =\ TRUE} argument
ensures that, from the full complement of such possible combinations,
every one is included as a statistic. By default,
\texttt{levels\ =\ TRUE} whereas \texttt{levels2\ =\ -1}, since we
frequently want to exclude at least one possible mixing combination to
avoid collinearity in a model that also includes the \texttt{edges}
term.

We may now use \texttt{levels2} in conjunction with \texttt{blocks} to
select exactly which of the \texttt{nodemix} combinations should be
constrained as fixed to reproduce the examples of \secref{sec:Dyads}.
First, we fix all dyads where the \texttt{group} values do not match:

\begin{Shaded}
\begin{Highlighting}[]
\KeywordTok{coef}\NormalTok{(}\KeywordTok{ergm}\NormalTok{(ncole }\OperatorTok{\textasciitilde{}}\StringTok{ }\NormalTok{edges, }\DataTypeTok{constraints =} \OperatorTok{\textasciitilde{}}\KeywordTok{blocks}\NormalTok{(}\StringTok{"Semester"}\NormalTok{, }\DataTypeTok{levels2 =} \KeywordTok{c}\NormalTok{(}\DecValTok{1}\NormalTok{, }\DecValTok{4}\NormalTok{))))}
\end{Highlighting}
\end{Shaded}

\begin{verbatim}
##     edges 
## -4.276666
\end{verbatim}

Second, we fix the dyads where \texttt{group} values do match:

\begin{Shaded}
\begin{Highlighting}[]
\KeywordTok{coef}\NormalTok{(}\KeywordTok{ergm}\NormalTok{(ncole }\OperatorTok{\textasciitilde{}}\StringTok{ }\NormalTok{edges, }\DataTypeTok{constraints =} \OperatorTok{\textasciitilde{}}\KeywordTok{blocks}\NormalTok{(}\StringTok{"Semester"}\NormalTok{, }\DataTypeTok{levels2 =} \KeywordTok{c}\NormalTok{(}\DecValTok{2}\NormalTok{, }\DecValTok{3}\NormalTok{))))}
\end{Highlighting}
\end{Shaded}

\begin{verbatim}
##     edges 
## -2.984404
\end{verbatim}

Additional examples using \texttt{levels2}, among other nodal attribute
features, are contained in the \texttt{nodal\_attributes} vignette
within the \pkg{ergm} package.

\hypertarget{additional-constraints}{%
\subsection{Additional constraints}\label{additional-constraints}}

Multiple different constraints on the sample space of possible networks,
as defined by the values of certain network statistics, may be
implemented beyond those discussed already in this section. The
\texttt{bd} constraint, for instance, may be used to enforce a maximum
allowable degree for any node, via the \texttt{maxout} argument. A
comprehensive list of available constraints is available via
\texttt{?\ ergmConstraint}. The handling of various constraints by MCMC
proposals in the \pkg{ergm} package is addressed in Krivitsky et al.
(2022).

\hypertarget{modeling-networks-with-valued-edges}{%
\section{Modeling Networks with Valued
Edges}\label{modeling-networks-with-valued-edges}}

\label{sec:Valued}

Starting with version 3.1, the \pkg{ergm} package can handle some types
of networks whose ties are not merely binary, indicating presence or
absence, but may have nonzero values other than unity. For example, the
value of a tie might represent a count, such as the number of times a
particular relationship has occurred; or it might represent an ordinal
variable, if node \(i\) ranks a subset of its neighbors. Valued ties can
increase complexity relative to binary ties in, for example, specifying
the model and ensuring that the chosen statistics are meaningful for the
types of edge values being modeled. Whether the scale of measurement of
tie values is ordinal, interval, or ratio, it becomes necessary to
specify the distribution of these values and to create functions to
aggregate these values into ERGM statistics.

In the \texttt{ergm()}, \texttt{simulate()}, and \texttt{summary()}
functions, the valued mode is typically activated by passing a
\texttt{response=} argument, giving the name of the edge attribute
containing the value of the response. Non-edges are assumed to have
value 0. The argument may also be a formula whose right-hand side is an
expression in terms of the edge attributes that evaluates to the
response value and whose left-hand side, if present, gives the name to
be used. If it evaluates to a logical (\texttt{TRUE}/\texttt{FALSE})
value (e.g.,
\texttt{response=threeContacts\textasciitilde{}(contacts\textgreater{}=3)}),
a binary ERGM is used.

\hypertarget{reference-specification}{%
\subsection{Reference specification}\label{reference-specification}}

\label{sec:ReferenceSpecification}

Krivitsky (2012) pointed out that sufficient statistics alone do not
suffice to specify an ERGM on a network whose relations are valued.
Consider a simple ERGM of the form \begin{equation}\label{SumStatistic}
  \Prob(\Yy;\curvpark[])\propto\h(\y)\myexp{\curvpark[] \sum_{\ijdysY}\yij},
\end{equation} where \(\yij\in\en\{\}{0,1,\dotsc}\) is an unbounded
count. If \(h(\y)\) is any constant, then
\(\Yij\iid \Geometric[p=1-\myexp{\curvpark[]}]\). On the other hand, if
\(h(\y)=1/\prod_{\ijdysY}\yij!\), then
\(\Yij\iid \Poisson[\mu=\myexp{\curvpark}]\). For this reason, Krivitsky
(2012) called a distribution with \(h(\y)=1\) and a sample space of
nonnegative integers a \emph{Geometric-reference ERGM} and one with
\(h(\y)=1/\prod_{\ijdysY}\yij!\) a \emph{Poisson-reference ERGM}.

For \texttt{ergm()}, \texttt{simulate()}, and other functions, reference
distributions are specified with a \texttt{reference=} argument, which
is a one-sided formula with one term. The \pkg{ergm} package allows
\texttt{Unif(a,b)} and \texttt{DiscUnif(a,b)} references, specifying
\(h(\y)=1\), the former on a dyad space \(\yij\in [a,b]\), the latter on
\(\yij\in\en\{\}{a,a+1,\dotsc,b}\). A companion package,
\texttt{ergm.count}, allows the additional references \texttt{Poisson}
and \texttt{Geometric}, described above, as well as
\texttt{Binomial(trials)} for
\(h(\y)=\prod_{\ijdysY} \binom{n_{\text{trials}}}{\yij}\) in the case
\(\yij\in\en\{\}{0,\dotsc,n _{\text{trials}}}\). For rank-order
relational data, a \texttt{CompleteOrder} reference distribution is
implemented in the \texttt{ergm.rank} package for situations where
rankings are compete. Where ties are permitted, \texttt{DiscUnif()} can
be used as a reference. See Krivitsky \& Butts (2019) for further
details on both the \pkg{ergm.count} and \pkg{ergm.rank} packages, and
their vignettes.

Reference distributions are explained in more detail in Section 3 of
Krivitsky \& Butts (2019). This reference also illustrates how the
\pkg{network} package may be used to visualize some kinds of valued
networks (Section 2) and even how the \pkg{latentnet} package can handle
latent-space models with valued ties (Section 4). Online help on the
reference distributions that are implemented by all packages currently
loaded in an \proglang{R} session can be obtained by typing
\texttt{help("ergm-references")}.

\hypertarget{dyad-independent-statistics}{%
\subsection{Dyad-Independent
statistics}\label{dyad-independent-statistics}}

\label{sec:DyadIndependent}

As in \secref{sec:NetworkFilters}, a component of the vector
\(\genstats(\y)\) is called a dyad-independent statistic if, when one
builds an ERGM using it as the \emph{only} model statistic, the joint
distribution \eqref{ergm} of the network factors into the product of its
marginal dyad distributions. That is, the univariate version of equation
\eqref{ergm} may be written \begin{equation}\label{DyadIndependentTerm}
\Prob(\Yy;\curvparel)
=\prod_{\ijdysY} \Prob( \Yij = \yij)
= \frac{\h(\y)}
{\normc_{\h,\cnmapel,\genstatel}(\curvparel,\netsY)}
\prod_{\ijdysY}
\exp\en\{\}{\cnmapel(\curvparel) \genstatel_{\ij}(\y)}
\end{equation} for \(\ynetsY\) and for some appropriately chosen
\(\genstatel_{\ij}(\y)\). Equation \eqref{SumStatistic} shows that the
sum of the values \(\yij\), which implies \(\genstatel_{\ij}(\y)=\yij\),
is one such example. Another example is the sum of the nonzero
indicators that arises if we define
\(\genstatel_{\ij}(\y)=\indf{\yij>0}\). Each of these basic
dyad-independent statistics is implemented in \pkg{ergm}:

\begin{description}
\item[\texttt{sum(pow=1)} \emph{Sum of edge values}]
This is simply the summation of edge values. For most valued ERGMs, this
is the natural intercept term. In particular, for reference
distributions such as \texttt{Poisson} and \texttt{Binomial}, using this
term produces intercept effects of Poisson log-linear and binomial
logistic regressions, respectively. Optionally, the dyad values can be
raised to a power before being summed.
\item[\texttt{nonzero} \emph{Number of nonzero edge values}]
This term counts nonzero edge values. It can be used to model
zero-inflation that is common in networks: It is often the case that a
network is sparse but has edges with relatively high weights when they
are present.
\end{description}

Binary ERGM statistics cannot be used directly for valued networks nor
vice versa, but most dyad-independent binary ERGM statistics have been
generalized by imposing a covariate on one of the two above forms. They
have the same arguments as their binary ERGM counterparts, with an
additional argument: \texttt{form}, which has two possible values:
\texttt{"sum"} (the default) and \texttt{"nonzero"}. The former creates
a statistic of the form \(\sum_{\ijdysY} x_{i,j} \yij\), where \(\yij\)
is the value of dyad \((i,j)\) and \(x_{i,j}\) is the term's covariate
associated with it. The latter computes a sum of indicator variables,
one for each dyad, indicating whether the corresponding edge has a
nonzero value. When \texttt{form="sum"} is used, typically a GLM-like
effect results, whereas \texttt{form="nonzero"} can be used to model
sparsity effects. (Krivitsky, 2012) Krivitsky \& Butts (2019) gives an
example of the \texttt{form=} argument with the \texttt{nodematch} term.

Other terms that control a dyad's distribution are
\texttt{atleast(threshold\ =\ 0)}, \texttt{atmost(threshold\ =\ 0)},
\texttt{equalto(value\ =\ 0,\ tolerance\ =\ 0)},
\texttt{greaterthan(threshold\ =\ 0)},
\texttt{ininterval(lower\ =\ -Inf,\ upper\ =\ +Inf,\ open\ =\ c(TRUE,\ TRUE))},
and \texttt{smallerthan(threshold\ =\ 0)}. Each of these terms counts
the dyad values that satisfy the criterion identified by its name.

\hypertarget{mutuality}{%
\subsection{Mutuality}\label{mutuality}}

The binary \texttt{mutuality} term in \texttt{ergm} counts the number of
pairs of mutual ties. Its valued counterpart is
\texttt{mutuality(form)}, which permits the following values of
\texttt{form}. For each of these, a higher coefficient will tend to
increase the similarity of reciprocating dyad values.

\begin{description}
\item[\texttt{"product"} \emph{Sum of products of reciprocating edge
values}]
This is the most direct generalization. However, for a
\texttt{Poisson}-reference ERGM in particular, a positive coefficient on
this term produces an infinite normalizing constant and therefore lies
outside the parameter space.
\item[\texttt{"geometric"} \emph{Sum of geometric mean of reciprocating
edge values}]
This form solves the \texttt{product} form's problem by taking a square
root of the product. It can be viewed as the uncentered covariance of
variance-stabilized counts.
\item[\texttt{"min"} \emph{Minimum of reciprocating edge values}]
This effect is, perhaps, the easiest to interpret, at the cost of
statistical power.
\item[\texttt{"nabsdiff"} \emph{Absolute difference of reciprocating
edge values}]
This effect is more symmetrical than \texttt{min}.
\end{description}

We refer the reader to Krivitsky (2012) for a further discussion of the
effects.

\hypertarget{actor-heterogeneity}{%
\subsection{Actor heterogeneity}\label{actor-heterogeneity}}

Different actors may have different overall propensities to interact.
This has been modeled using random effects, as in the \(p_2\) model, and
using degeneracy-prone terms like \(k\)-star counts. For valued ERGMs,
the following term, also introduced by Krivitsky (2012) and discussed in
more detail there, models actor heterogeneity:

\begin{description}
\item[\texttt{nodesqrtcovar(center,transform)} \emph{Covariance between
\(\yij\) incident on same actor}]
The default \texttt{transform="sqrt"} will take a square root of dyad
values before calculating, and the default \texttt{center=TRUE} will
center the transformed values around their global mean, gaining
stability at the cost of locality.
\end{description}

\hypertarget{triadic-effects}{%
\subsection{Triadic effects}\label{triadic-effects}}

To generalize the notion of triadic closure, \texttt{ergm} implements
very flexible \texttt{transitiveweights(twopath,\ combine,\ affect)} and
similar \texttt{cyclicalweights} statistics. The transitive weight
statistic has the general form
\[\genstat{$\boldsymbol{v}$}(\y)=\sum_{\ijdysY}v_{\text{affect}}
\left(\yij,v_{\text{combine}}\left(v_{\text{2-path}}
(\y_{i,k},\y_{k,j})_{k\in N\setsub \{i,j\}}\right)\right),\] which can
be customized by varying three functions:

\begin{description}
\item[\(v_{\text{2-path}}\)]
Given \(\y_{i,k}\) and \(\y_{k,j}\), what is the strength of the
two-path they form? The options are \texttt{"min"}, to take the minimum
of the two-path's constituent values, and \texttt{"geomean"}, to take
their geometric mean, gaining statistical power at a greater risk of
model instability.
\item[\(v_{\text{combine}}\)]
Given the strengths of the two-paths \(\y_{i\to k\to j}\) for all
\(k\ne i,j\), what is the combined strength of these two-paths between
\(i\) and \(j\)? The choices are \texttt{"max"}, for the strength of the
strongest of the two-paths---analogous to \texttt{transitiveties} or
\texttt{gwesp(0)} binary ERGM effects---and \texttt{"sum"}, the sum of
the path strengths. The latter choice is better able to detect effects
but is more subject to degeneracy; it is analogous to
\texttt{triangles}.
\item[\(v_{\text{affect}}\)]
Given the combined strength of the two-paths between \(i\) and \(j\),
how should they affect \(\Yij\)? The choices are \texttt{"min"}, the
minimum of the combined strength and the focus two-path, and
\texttt{"geomean"}, again more able to detect effects but more likely to
cause degeneracy.
\end{description}

Usage of the \texttt{transitiveweights} and \texttt{cyclicalweights}
terms is illustrated in Section 3.1 of Krivitsky \& Butts (2019).

\hypertarget{using-binary-ergm-terms-in-valued-ergms}{%
\subsection{Using binary ERGM terms in valued
ERGMs}\label{using-binary-ergm-terms-in-valued-ergms}}

\pkg{ergm} also allows general binary terms to be passed to valued
models. The mechanism that allows this is the term operator
\texttt{B(formula,\ form)}, which is further described in the \pkg{ergm}
online help under \texttt{help("B-ergmTerm")} or the shorthand
\texttt{ergmTerm?B}. Here, \texttt{formula=} is a one-sided formula
whose right hand side contains the binary \texttt{ergm} terms to be
used. Allowable values of the \texttt{form} argument are
\texttt{form="sum"} and \texttt{form="nonzero"}, which have the effects
described in \secref{sec:DyadIndependent}, with \texttt{form="sum"} only
valid for dyad-independent \texttt{formula=} terms; or a one-sided
formula may be passed to \texttt{form=}, containing one \emph{valued}
\texttt{ergm} term, with the following properties:

\begin{itemize}
\tightlist
\item
  dyadic independence;
\item
  dyadwise contribution of either 0 or 1;
\item
  dyadwise contribution of 0 for a 0-valued dyad.
\end{itemize}

That is, it must be expressible as
\[\genstatel(y) = \sum_{\ijdysY} \genstatel _
{i,j}(\yij),\] where for all \(i\), \(j\), and \(\y\),
\(\genstatel _ {i,j}(\yij)\in\en\{\}{0,1}\) and
\(\genstatel_{i,j}(0)\equiv 0\). Such terms include \texttt{nonzero},
\texttt{ininterval()}, \texttt{atleast()}, \texttt{atmost()},
\texttt{greaterthan()}, \texttt{lessthan()}, and \texttt{equalto()}. The
operator will then construct a binary network \(\y\binary\) such that
\(\yij\binary=1\) if and only if \(\genstatel _ {i,j}(\yij) = 1\), and
evaluate the binary terms in \texttt{formula=} on it.

\hypertarget{modeling-ordinal-values-using-binary-term-operators}{%
\subsection{Modeling Ordinal Values Using Binary Term
Operators}\label{modeling-ordinal-values-using-binary-term-operators}}

To illustrate the use of binary ergm terms on a valued network as
described above, we construct an example that uses the \texttt{B} (for
``binary'') operator. The code snippet below gives an example of a
valued ergm that uses the \texttt{DiscUnif}, or discrete uniform,
reference distribution, which is included in the \pkg{ergm} package
itself; that is, there is no need to load the \pkg{ergm.count} or
\pkg{ergm.rank} packages to run the following example. The example fits
a multinomial logistic regression model that assumes that the edge
values independent of one another and take ordinal values that have the
same interpretation for each dyad. (In general, rating and ranking data
may not allow edge values to be compared across egos (Krivitsky \&
Butts, 2017); the \pkg{ergm.rank} package contains terms that remain
valid in this more complex setting.) Models for independently observed
ordinal random variables have a long history in the statistical
literature; relevant references specific to network models include
Robins et al. (1999) and, in a Bayesian framework, Caimo \& Gollini
(2020).

First, we build a valued network by pooling the three binary friendship
nomination networks due to Sampson (1968), exactly as in Section 2.1 of
Krivitsky \& Butts (2019).

\begin{Shaded}
\begin{Highlighting}[]
\KeywordTok{data}\NormalTok{(samplk)}
\CommentTok{\# Create a sociomatrix totaling the nominations.}
\NormalTok{samplk.tot.m \textless{}{-}}\StringTok{ }\KeywordTok{as.matrix}\NormalTok{(samplk1) }\OperatorTok{+}\StringTok{ }\KeywordTok{as.matrix}\NormalTok{(samplk2) }\OperatorTok{+}\StringTok{ }\KeywordTok{as.matrix}\NormalTok{(samplk3)}
\NormalTok{samplk.tot \textless{}{-}}\StringTok{ }\KeywordTok{as.network}\NormalTok{(samplk.tot.m, }\DataTypeTok{directed=}\OtherTok{TRUE}\NormalTok{, }\DataTypeTok{matrix.type=}\StringTok{"a"}\NormalTok{,}
                         \DataTypeTok{ignore.eval=}\OtherTok{FALSE}\NormalTok{, }\DataTypeTok{names.eval=}\StringTok{"nominations"}\NormalTok{)}
\end{Highlighting}
\end{Shaded}

We will use the \texttt{B} operator to construct new statistics
consisting of the number of edges with value \(k\) or higher, where
\(k\) is 1, 2, or 3.

\begin{Shaded}
\begin{Highlighting}[]
\KeywordTok{summary}\NormalTok{(samplk.tot }\OperatorTok{\textasciitilde{}}\StringTok{ }\KeywordTok{B}\NormalTok{(}\OperatorTok{\textasciitilde{}}\NormalTok{edges, }\OperatorTok{\textasciitilde{}}\KeywordTok{atleast}\NormalTok{(}\DecValTok{1}\NormalTok{)) }\OperatorTok{+}\StringTok{ }\KeywordTok{B}\NormalTok{(}\OperatorTok{\textasciitilde{}}\NormalTok{edges, }\OperatorTok{\textasciitilde{}}\KeywordTok{atleast}\NormalTok{(}\DecValTok{2}\NormalTok{))}
                   \OperatorTok{+}\StringTok{ }\KeywordTok{B}\NormalTok{(}\OperatorTok{\textasciitilde{}}\NormalTok{edges, }\OperatorTok{\textasciitilde{}}\KeywordTok{atleast}\NormalTok{(}\DecValTok{3}\NormalTok{)), }\DataTypeTok{response =} \StringTok{"nominations"}\NormalTok{)}
\end{Highlighting}
\end{Shaded}

\begin{verbatim}
## B(atleast(1))~edges B(atleast(2))~edges B(atleast(3))~edges 
##                  88                  50                  30
\end{verbatim}

Since there are \(18\times17\), or 306, possible edges, the summary
statistics above tell us that the valued network we have constructed has
30 edges with value 3, 20 with value 2, 38 with value 1, and the
remaining 218 with value 0. The ERGM with these statistics has
independent edges, where the probabilities an edge takes the values 0,
1, 2, or 3 are given by \(1/D\), \(\exp\{\curvpark[1]\}/D\),
\(\exp\{\curvpark[1]+\curvpark[2]\}/D\), and
\(\exp\{\curvpark[1]+\curvpark[2]+\curvpark[3]\}/D\), respectively,
where \[
D = 1 + \exp\{\curvpark[1]\} + \exp\{\curvpark[1]+\curvpark[2]\} + \exp\{\curvpark[1]+\curvpark[2]+\curvpark[3]\}.
\] We may verify that \texttt{ergm}'s stochastic fitting algorithm
obtains MLEs very close to the exact values:

\begin{Shaded}
\begin{Highlighting}[]
\NormalTok{mod \textless{}{-}}\StringTok{ }\KeywordTok{ergm}\NormalTok{(samplk.tot }\OperatorTok{\textasciitilde{}}\StringTok{ }\KeywordTok{B}\NormalTok{(}\OperatorTok{\textasciitilde{}}\NormalTok{edges, }\OperatorTok{\textasciitilde{}}\KeywordTok{atleast}\NormalTok{(}\DecValTok{1}\NormalTok{)) }\OperatorTok{+}\StringTok{ }\KeywordTok{B}\NormalTok{(}\OperatorTok{\textasciitilde{}}\NormalTok{edges, }\OperatorTok{\textasciitilde{}}\KeywordTok{atleast}\NormalTok{(}\DecValTok{2}\NormalTok{))}
          \OperatorTok{+}\StringTok{ }\KeywordTok{B}\NormalTok{(}\OperatorTok{\textasciitilde{}}\NormalTok{edges, }\OperatorTok{\textasciitilde{}}\KeywordTok{atleast}\NormalTok{(}\DecValTok{3}\NormalTok{)), }\DataTypeTok{response =} \StringTok{"nominations"}\NormalTok{,}
          \DataTypeTok{reference =} \OperatorTok{\textasciitilde{}}\KeywordTok{DiscUnif}\NormalTok{(}\DecValTok{0}\NormalTok{,}\DecValTok{3}\NormalTok{), }\DataTypeTok{control =} \KeywordTok{snctrl}\NormalTok{(}\DataTypeTok{seed =} \DecValTok{123}\NormalTok{))}
\KeywordTok{coef}\NormalTok{(mod) }\CommentTok{\# Approximate MLEs for theta1, theta2, and theta3}
\end{Highlighting}
\end{Shaded}

\begin{verbatim}
## B(atleast(1))~edges B(atleast(2))~edges B(atleast(3))~edges 
##          -1.7342318          -0.6602796           0.4090445
\end{verbatim}

\begin{Shaded}
\begin{Highlighting}[]
\NormalTok{true \textless{}{-}}\StringTok{ }\KeywordTok{c}\NormalTok{(}\DataTypeTok{EdgeVal0=}\DecValTok{218}\NormalTok{, }\DataTypeTok{EdgeVal1=}\DecValTok{38}\NormalTok{, }\DataTypeTok{EdgeVal2=}\DecValTok{20}\NormalTok{, }\DataTypeTok{EdgeVal3=}\DecValTok{30}\NormalTok{)}
\NormalTok{est \textless{}{-}}\StringTok{ }\KeywordTok{c}\NormalTok{(}\DecValTok{1}\NormalTok{, }\KeywordTok{exp}\NormalTok{(}\KeywordTok{cumsum}\NormalTok{(}\KeywordTok{coef}\NormalTok{(mod))), }\DataTypeTok{use.names=}\OtherTok{FALSE}\NormalTok{)}
\KeywordTok{rbind}\NormalTok{(}\DataTypeTok{True\_Proportions =}\NormalTok{ true }\OperatorTok{/}\StringTok{ }\KeywordTok{sum}\NormalTok{(true), }\DataTypeTok{Estimated\_Proportions =}\NormalTok{ est }\OperatorTok{/}\StringTok{ }\KeywordTok{sum}\NormalTok{(est))}
\end{Highlighting}
\end{Shaded}

\begin{verbatim}
##                        EdgeVal0 EdgeVal1   EdgeVal2   EdgeVal3
## True_Proportions      0.7124183 0.124183 0.06535948 0.09803922
## Estimated_Proportions 0.7117086 0.125642 0.06492009 0.09772932
\end{verbatim}

This example could have used the \texttt{equalto} terms in place of all
the \texttt{atleast} terms above. Then, the estimated proportions would
have been proportional to 1, \(\exp\{\curvpark[1]\}\),
\(\exp\{\curvpark[2]\}\), and \(\exp\{\curvpark[3]\}\) instead of 1,
\(\exp\{\curvpark[1]\}\), \(\exp\{\curvpark[1]+\curvpark[2]\}\), and
\(\exp\{\curvpark[1]+\curvpark[2]+\curvpark[3]\}\). Such a model does
not assume ordinality of the edge values, so it could be used for a
multinomial logit model in which the edges take categorical non-ordered
values.

\hypertarget{estimation-in-the-presence-of-missing-edge-data}{%
\section{Estimation in the presence of missing edge
data}\label{estimation-in-the-presence-of-missing-edge-data}}

\label{sec:Missing}

It is quite common that network data are incomplete in various ways. The
\pkg{ergm} package includes the capability to handle missing edge data,
whereas other types of missingness such as missing nodal information are
not addressed. Handcock \& Gile (2010) formulated a framework for
modeling networks with missing ties and expressed the log-likelihood as
\begin{equation}\label{eq:loglikelihood}
\llik(\curvpar)=\log \Prob(\Y\in\netsY(\yobs);\curvpar)=\log\sum_{\ypnetsY(\yobs)}\Prob(\Y=\y';\curvpar),
\end{equation} where \(\netsY(\yobs)\) is defined as the set of networks
whose partial observation could have produced \(\yobs\): essentially,
all of the ways to impute the missing ties in \(\yobs\). (When there are
no missing ties in \(\yobs\), \(\netsY(\yobs)\) contains only
\(\yobs\).) They then proposed to maximize this likelihood by taking
advantage of the fact that, if
\[\cY{\netsY'}(\curvpar)\defeq\sum_{\y'\in\netsY'}\h(\y')\exp\en\{\}{\natpar\t\genstats(\y')},\]
the log-likelihood can be expressed as
\(\llik(\curvpar)=\log \cY{\netsY(\yobs)}(\curvpar) - \log \cy(\curvpar),\)
resulting in the score equation
\[\grad_{\curvpar}\llik(\mle)=\dnatmle\t\en[]{\EY{\netsY(\yobs)}\en\{\}{\genstats(\Y);\mle}-\Ey\en\{\}{\genstats(\Y);{\mle}}}=\0,\]
with MCMLE approximation also possible for the first term by fixing a
particular \(\guess\) and drawing a sample from
\(\DY{\netsY(\yobs)}(\guess)\) as explained in Section 3 of Krivitsky et
al. (2022).

The \pkg{ergm} package invokes the above approach automatically when a
network has missing edge variables. The simplest way to encode a missing
edge is to set its value to \texttt{NA}. The \pkg{network} package
natively supports missing edge variables coded in this way, and
\texttt{network} objects with missingness are thus handled without
additional intervention. \pkg{ergm}'s methods for assessing goodness of
fit of a model by comparing observed values of certain network
statistics to the distribution of their simulated values under the model
(Hunter, Goodreau, et al., 2008, p. @HuHa08e) have also been adapted to
missing edge data: the (unavailable) observed values of the statistics
of interest \(t(\y)\) are replaced by their conditional expectations
\(\EY{\netsY(\yobs)}\en\{\}{t(\Y);\mle}\).

Here we fit a simple model with edges, mutuality (reciprocated dyads),
transitive ties, and cyclical ties to the Sampson Monks dataset depicted
in \figref{fig:sampson}. For the sake of comparison, we first fit the
model assuming no missing edge data, which may be quickly verified using
the output of the \texttt{print(samplike)} command:

\begin{Shaded}
\begin{Highlighting}[]
\KeywordTok{print}\NormalTok{(samplike)}
\end{Highlighting}
\end{Shaded}

\begin{verbatim}
##  Network attributes:
##   vertices = 18 
##   directed = TRUE 
##   hyper = FALSE 
##   loops = FALSE 
##   multiple = FALSE 
##   total edges= 88 
##     missing edges= 0 
##     non-missing edges= 88 
## 
##  Vertex attribute names: 
##     cloisterville group vertex.names 
## 
##  Edge attribute names: 
##     nominations
\end{verbatim}

\begin{Shaded}
\begin{Highlighting}[]
\KeywordTok{summary}\NormalTok{(full.fit \textless{}{-}}\StringTok{ }\KeywordTok{ergm}\NormalTok{(samplike }\OperatorTok{\textasciitilde{}}\StringTok{ }\NormalTok{edges }\OperatorTok{+}\StringTok{ }\NormalTok{mutual }\OperatorTok{+}\StringTok{ }\NormalTok{transitiveties }\OperatorTok{+}\StringTok{ }\NormalTok{cyclicalties,}
                         \DataTypeTok{eval.loglik=}\OtherTok{TRUE}\NormalTok{), }\DataTypeTok{control =} \KeywordTok{snctrl}\NormalTok{(}\DataTypeTok{seed =} \DecValTok{321}\NormalTok{))}
\end{Highlighting}
\end{Shaded}

\begin{verbatim}
## Call:
## ergm(formula = samplike ~ edges + mutual + transitiveties + cyclicalties, 
##     eval.loglik = TRUE)
## 
## Monte Carlo Maximum Likelihood Results:
## 
##                Estimate Std. Error MCMC % z value Pr(>|z|)    
## edges           -1.9436     0.3542      0  -5.488   <1e-04 ***
## mutual           2.5066     0.4551      0   5.507   <1e-04 ***
## transitiveties   0.5499     0.2880      0   1.909   0.0562 .  
## cyclicalties    -0.4582     0.2393      0  -1.915   0.0555 .  
## ---
## Signif. codes:  0 '***' 0.001 '**' 0.01 '*' 0.05 '.' 0.1 ' ' 1
## 
##      Null Deviance: 424.2  on 306  degrees of freedom
##  Residual Deviance: 329.0  on 302  degrees of freedom
##  
## AIC: 337  BIC: 351.9  (Smaller is better. MC Std. Err. = 0.5904)
\end{verbatim}

Now, suppose that Monk \#1 (John Bosco) refused to respond during all
three waves, rendering his replies missing:

\begin{Shaded}
\begin{Highlighting}[]
\NormalTok{samplike1 \textless{}{-}}\StringTok{ }\NormalTok{samplike}
\NormalTok{samplike1[}\DecValTok{1}\NormalTok{, ] \textless{}{-}}\StringTok{ }\OtherTok{NA}
\KeywordTok{print}\NormalTok{(samplike1)}
\end{Highlighting}
\end{Shaded}

\begin{verbatim}
##  Network attributes:
##   vertices = 18 
##   directed = TRUE 
##   hyper = FALSE 
##   loops = FALSE 
##   multiple = FALSE 
##   total edges= 99 
##     missing edges= 17 
##     non-missing edges= 82 
## 
##  Vertex attribute names: 
##     cloisterville group vertex.names 
## 
##  Edge attribute names: 
##     nominations
\end{verbatim}

If we pass this modified object to \texttt{ergm}, it will automatically
calculate the MLE under the assumption that the monk's refusal is
unrelated to his choice of relations, i.e., that the data are ignorably
missing with respect to the specified model:

\begin{Shaded}
\begin{Highlighting}[]
\KeywordTok{summary}\NormalTok{(m1.fit \textless{}{-}}\StringTok{ }\KeywordTok{ergm}\NormalTok{(samplike1}\OperatorTok{\textasciitilde{}}\NormalTok{edges}\OperatorTok{+}\NormalTok{mutual}\OperatorTok{+}\NormalTok{transitiveties}\OperatorTok{+}\NormalTok{cyclicalties,}
                       \DataTypeTok{eval.loglik=}\OtherTok{TRUE}\NormalTok{), }\DataTypeTok{control =} \KeywordTok{snctrl}\NormalTok{(}\DataTypeTok{seed =} \DecValTok{321}\NormalTok{))}
\end{Highlighting}
\end{Shaded}

\begin{verbatim}
## Call:
## ergm(formula = samplike1 ~ edges + mutual + transitiveties + 
##     cyclicalties, eval.loglik = TRUE)
## 
## Monte Carlo Maximum Likelihood Results:
## 
##                Estimate Std. Error MCMC % z value Pr(>|z|)    
## edges           -2.0142     0.3839      0  -5.246   <1e-04 ***
## mutual           2.4206     0.4677      0   5.175   <1e-04 ***
## transitiveties   0.4726     0.4001      0   1.181    0.238    
## cyclicalties    -0.3013     0.3455      0  -0.872    0.383    
## ---
## Signif. codes:  0 '***' 0.001 '**' 0.01 '*' 0.05 '.' 0.1 ' ' 1
## 
##      Null Deviance: 400.6  on 289  degrees of freedom
##  Residual Deviance: 314.5  on 285  degrees of freedom
##  
## AIC: 322.5  BIC: 337.2  (Smaller is better. MC Std. Err. = 0.5524)
\end{verbatim}

The degrees of freedom associated with the missing data fit have
decreased because unobserved dyads do not carry information. For details
regarding the ignorability assumption for edge variables, see Handcock
\& Gile (2010).

The estimation approach above can be extended to other types of
incomplete network observation. Karwa et al. (2017) applied it to fit
arbitrary ERGMs to networks whose dyad values had been stochastically
perturbed---ties added and removed at random, with known
probabilities---in order to preserve privacy. Another use case is
multiple imputation for networks with missing data, in which multiple
random versions of the full network are constructed by randomly
inserting values for unobserved dyads according to probabilities that
are determined based on, say, some type of logistic regression model.
These mechanisms may be invoked by passing an \texttt{obs.constraints}
formula, specifying how the network of interest was observed. Of
particular interest are the following constraints:

\begin{description}
\item[\texttt{observed}]
restricts the proposal to changing only those dyads that are recorded as
missing.
\item[\texttt{egocentric(attr\ =\ NULL,\ direction\ =\ c("both","out","in"))}]
restricts the proposal to changing only those dyads that would not be
observed in an egocentric sample. That is, dyads cannot be modified that
are incident on vertices for which attribute specification \texttt{attr}
has value \texttt{TRUE} or, if \texttt{attr} is \texttt{NULL}, the
vertex attribute \texttt{"na"} has value \texttt{FALSE}. For directed
networks, \texttt{direction=="out"} only preserves the out-dyads of
those actors, and \texttt{direction=="in"} preserves their in-dyads.
\item[\texttt{dyadnoise(p01,p10)}]
Unlike the others, this is a soft constraint to adjust the sampled
distribution for dyad-level noise with known perturbation probabilities,
which can arise in a variety of contexts (Karwa et al., 2017). It is
assumed that the observed LHS network is a noisy observation of some
unobserved true network, with \texttt{p01} giving the dyadwise
probability of erroneously observing a tie where the true network had a
non-tie and \texttt{p10} giving the dyadwise probability of erroneously
observing a nontie where the true network had a tie. \texttt{p01} and
\texttt{p10} can be either both be scalars or both be adjacency matrices
of the same dimension as that of the LHS network giving these
probabilities.
\end{description}

We may use the \texttt{obs.constraints} argument to re-fit the model
above:

\begin{Shaded}
\begin{Highlighting}[]
\NormalTok{samplike2 \textless{}{-}}\StringTok{ }\NormalTok{samplike}
\NormalTok{samplike2[}\DecValTok{1}\NormalTok{,] \textless{}{-}}\StringTok{ }\DecValTok{0}
\NormalTok{samplike2 }\OperatorTok{\%v\%}\StringTok{ "refused"}\NormalTok{ \textless{}{-}}\StringTok{ }\KeywordTok{rep}\NormalTok{(}\KeywordTok{c}\NormalTok{(}\OtherTok{TRUE}\NormalTok{,}\OtherTok{FALSE}\NormalTok{),}\KeywordTok{c}\NormalTok{(}\DecValTok{1}\NormalTok{,}\DecValTok{17}\NormalTok{))}
\NormalTok{samplike2 }\CommentTok{\# same as print(samplike2)}
\end{Highlighting}
\end{Shaded}

\begin{verbatim}
##  Network attributes:
##   vertices = 18 
##   directed = TRUE 
##   hyper = FALSE 
##   loops = FALSE 
##   multiple = FALSE 
##   total edges= 82 
##     missing edges= 0 
##     non-missing edges= 82 
## 
##  Vertex attribute names: 
##     cloisterville group refused vertex.names 
## 
##  Edge attribute names: 
##     nominations
\end{verbatim}

\begin{Shaded}
\begin{Highlighting}[]
\KeywordTok{summary}\NormalTok{(m2.fit \textless{}{-}}\StringTok{ }\KeywordTok{ergm}\NormalTok{(samplike2 }\OperatorTok{\textasciitilde{}}\StringTok{ }\NormalTok{edges }\OperatorTok{+}\StringTok{ }\NormalTok{mutual }\OperatorTok{+}\StringTok{ }\NormalTok{transitiveties }\OperatorTok{+}\StringTok{ }\NormalTok{cyclicalties,}
                       \DataTypeTok{obs.constraints =} \OperatorTok{\textasciitilde{}}\StringTok{ }\KeywordTok{egocentric}\NormalTok{(}\OperatorTok{\textasciitilde{}!}\NormalTok{refused, }\StringTok{"out"}\NormalTok{),}
                       \DataTypeTok{control =} \KeywordTok{snctrl}\NormalTok{(}\DataTypeTok{seed =} \DecValTok{123}\NormalTok{)))}
\end{Highlighting}
\end{Shaded}

\begin{verbatim}
## Call:
## ergm(formula = samplike2 ~ edges + mutual + transitiveties + 
##     cyclicalties, obs.constraints = ~egocentric(~!refused, "out"), 
##     control = snctrl(seed = 123))
## 
## Monte Carlo Maximum Likelihood Results:
## 
##                Estimate Std. Error MCMC % z value Pr(>|z|)    
## edges           -2.0090     0.3867      0  -5.196   <1e-04 ***
## mutual           2.3890     0.4764      0   5.015   <1e-04 ***
## transitiveties   0.4401     0.4353      0   1.011    0.312    
## cyclicalties    -0.2660     0.3918      0  -0.679    0.497    
## ---
## Signif. codes:  0 '***' 0.001 '**' 0.01 '*' 0.05 '.' 0.1 ' ' 1
## 
##      Null Deviance: 400.6  on 289  degrees of freedom
##  Residual Deviance: 313.9  on 285  degrees of freedom
##  
## AIC: 321.9  BIC: 336.6  (Smaller is better. MC Std. Err. = 0.4579)
\end{verbatim}

Finally, since the observational process can be viewed as a part of the
network dataset, we may specify it using the \texttt{\%ergmlhs\%}
operation, giving a third way to fit the model above:

\begin{Shaded}
\begin{Highlighting}[]
\NormalTok{samplike2 }\OperatorTok{\%ergmlhs\%}\StringTok{ "obs.constraints"}\NormalTok{ \textless{}{-}}\StringTok{ }\ErrorTok{\textasciitilde{}}\KeywordTok{egocentric}\NormalTok{(}\OperatorTok{\textasciitilde{}!}\NormalTok{refused, }\StringTok{"out"}\NormalTok{)}
\KeywordTok{summary}\NormalTok{(m3.fit \textless{}{-}}\StringTok{ }\KeywordTok{ergm}\NormalTok{(samplike2 }\OperatorTok{\textasciitilde{}}\StringTok{ }\NormalTok{edges }\OperatorTok{+}\StringTok{ }\NormalTok{mutual }\OperatorTok{+}\StringTok{ }\NormalTok{transitiveties }\OperatorTok{+}\StringTok{ }\NormalTok{cyclicalties),}
        \DataTypeTok{control =} \KeywordTok{snctrl}\NormalTok{(}\DataTypeTok{seed =} \DecValTok{231}\NormalTok{))}
\end{Highlighting}
\end{Shaded}

\begin{verbatim}
## Call:
## ergm(formula = samplike2 ~ edges + mutual + transitiveties + 
##     cyclicalties)
## 
## Monte Carlo Maximum Likelihood Results:
## 
##                Estimate Std. Error MCMC % z value Pr(>|z|)    
## edges           -2.0540     0.3964      0  -5.182   <1e-04 ***
## mutual           2.4278     0.4918      0   4.937   <1e-04 ***
## transitiveties   0.5004     0.4377      0   1.143    0.253    
## cyclicalties    -0.3118     0.3949      0  -0.789    0.430    
## ---
## Signif. codes:  0 '***' 0.001 '**' 0.01 '*' 0.05 '.' 0.1 ' ' 1
## 
##      Null Deviance: 400.6  on 289  degrees of freedom
##  Residual Deviance: 311.8  on 285  degrees of freedom
##  
## AIC: 319.8  BIC: 334.4  (Smaller is better. MC Std. Err. = 0.6203)
\end{verbatim}

\hypertarget{other-enhancements}{%
\section{Other enhancements}\label{other-enhancements}}

We close this paper by highlighting a number of miscellaneous
enhancements to the \pkg{ergm} package since the Hunter, Handcock, et
al. (2008) article.

\hypertarget{exact-calculations-for-small-networks}{%
\subsection{Exact calculations for small
networks}\label{exact-calculations-for-small-networks}}

For small networks, it is possible to obtain full enumeration of all
possible network statistic vectors over the entire sample space of
possible networks. This enumeration enables exact calculations of such
quantities as the log-likelihood function, the MLE, or the normalizing
constant. If we consider only binary networks on an unconstrained sample
space, the total number of networks is \(2^{n(n-1)/2}\) for undirected
networks and \(2^{n(n-1)}\) for directed networks, which imposes a
practical limit of \(n=8\) nodes in the undirected case or \(n=6\) in
the directed case unless the user wants to compute for a long time, and
the functions described in this section return an error for larger
networks than these unless the \texttt{force=TRUE} option is invoked.

The \texttt{ergm.allstats} function, added to the \pkg{ergm} more than a
decade ago in version 2.4, performs an efficient, ``brute-force''
tabulation of all possible network statistic vectors for an arbitrary
ERGM by visiting every possible network. The \texttt{ergm.exact}
function uses \texttt{ergm.allstats} to calculate exact likelihood
values. Due to the computationally intractable normalizing constant
\(\cheg(\curvpar,\netsY)\) of Equation \eqref{ergm}, except in the case
of dyadic independence models, \texttt{ergm.exact} and
\texttt{ergm.allstats} may only be used for small networks. In a test,
the code below took about 254 times as long on a 9-node network as it
did on an 8-node network, which is not surprising because the 9-node
sample space has \(2^{36-28}\), or 256, times as many networks.

\begin{Shaded}
\begin{Highlighting}[]
\KeywordTok{system.time}\NormalTok{(\{}
\NormalTok{  EmptyNW \textless{}{-}}\StringTok{ }\KeywordTok{network.initialize}\NormalTok{(}\DecValTok{8}\NormalTok{, }\DataTypeTok{directed =} \OtherTok{FALSE}\NormalTok{) }\CommentTok{\# Replacing 8 by 9 takes much longer!}
\NormalTok{  a \textless{}{-}}\StringTok{ }\KeywordTok{ergm.allstats}\NormalTok{(EmptyNW }\OperatorTok{\textasciitilde{}}\StringTok{ }\NormalTok{edges }\OperatorTok{+}\StringTok{ }\NormalTok{triangle }\OperatorTok{+}\StringTok{ }\NormalTok{isolates }\OperatorTok{+}\StringTok{ }\KeywordTok{degree}\NormalTok{(}\DecValTok{4}\NormalTok{), }\DataTypeTok{force =} \OtherTok{TRUE}\NormalTok{)}
\NormalTok{\})}
\end{Highlighting}
\end{Shaded}

\begin{verbatim}
##    user  system elapsed 
##  72.319   0.003  72.352
\end{verbatim}

Naturally, many networks of interest are too large to utilize
\texttt{ergm.allstats} and \texttt{ergm.exact}. Yet calculations on
small networks can still provide useful test cases; for instance, see
Schmid \& Hunter (2020) or Vega Yon et al. (2021).

\hypertarget{estimation-based-only-on-sufficient-statistics}{%
\subsection{Estimation based only on sufficient
statistics}\label{estimation-based-only-on-sufficient-statistics}}

\label{sec:SufficientStatistics}

In exponential family parlance, \(\genstats(\yobs)\) is often called the
vector of sufficient statistics. Since the likelihood function of
Equation \eqref{eq:loglikelihood} depends on \(\yobs\) only via these
sufficient statistics, it is not actually necessary to observe \(\yobs\)
in order to calculate an MLE. The MLE-finding algorithm in \pkg{ergm}
exploits this fact by implementing the idea of Hummel et al. (2012) to
replace \(\genstats(\yobs)\) by a vector of statistics that is closer to
the sample mean generated by a current fixed, known parameter value.
Maximizing the resulting version of the loglikelihood function yields a
parameter value which may then be used to generate a new random sample
of networks, and the process is repeated to give a sequence of parameter
values approaching the desired MLE.

In some applications, such as when data are egocentrically sampled, it
is possible to observe or estimate the vector \(\genstats(\yobs)\) of
statistics that would in principle have been observed in the network,
even if other information about the network itself is absent. Estimation
may still proceed by passing a \texttt{target.stats} argument containing
a vector of network statistics. For example, we may reproduce (up to the
stochasticity of the fitting algorithm) the analysis of the
\texttt{full.fit} example in \secref{sec:Missing} by passing the vector
of statistics on the \texttt{samplike} network via \texttt{target.stats}
even though the network used in the \texttt{ergm} function call has no
edges at all:

\begin{Shaded}
\begin{Highlighting}[]
\CommentTok{\# Complete network statistics:}
\NormalTok{ts \textless{}{-}}\StringTok{ }\KeywordTok{summary}\NormalTok{(samplike }\OperatorTok{\textasciitilde{}}\StringTok{ }\NormalTok{edges }\OperatorTok{+}\StringTok{ }\NormalTok{mutual }\OperatorTok{+}\StringTok{ }\NormalTok{transitiveties }\OperatorTok{+}\StringTok{ }\NormalTok{cyclicalties)}
\NormalTok{emptynw \textless{}{-}}\StringTok{ }\KeywordTok{network.initialize}\NormalTok{(}\KeywordTok{network.size}\NormalTok{(samplike), }\DataTypeTok{directed =} \OtherTok{TRUE}\NormalTok{)}
\NormalTok{ts.fit \textless{}{-}}\StringTok{ }\KeywordTok{ergm}\NormalTok{(emptynw }\OperatorTok{\textasciitilde{}}\StringTok{ }\NormalTok{edges }\OperatorTok{+}\StringTok{ }\NormalTok{mutual }\OperatorTok{+}\StringTok{ }\NormalTok{transitiveties }\OperatorTok{+}\StringTok{ }\NormalTok{cyclicalties,}
               \DataTypeTok{target.stats =}\NormalTok{ ts, }\DataTypeTok{control =} \KeywordTok{snctrl}\NormalTok{(}\DataTypeTok{seed =} \DecValTok{123}\NormalTok{))}
\KeywordTok{rbind}\NormalTok{(}\KeywordTok{coef}\NormalTok{(full.fit), }\KeywordTok{coef}\NormalTok{(ts.fit))}
\end{Highlighting}
\end{Shaded}

\begin{verbatim}
##          edges   mutual transitiveties cyclicalties
## [1,] -1.943639 2.506586      0.5498717   -0.4582105
## [2,] -1.943678 2.498313      0.5192654   -0.4326818
\end{verbatim}

\hypertarget{predicting-individual-edge-probabilities}{%
\subsection{Predicting individual edge
probabilities}\label{predicting-individual-edge-probabilities}}

\label{sec:Predict}

The \texttt{predict} method, which may be called on either
\texttt{formula} or \texttt{ergm} objects, calculates model-predicted
conditional or unconditional tie probabilities for dyads in a binary
network. In the conditional case, \texttt{predict} simply uses the
output from the \texttt{ergmMPLE} function (see Krivitsky et al., 2022,
Section 3.1). In the unconditional case, even for dyadic independence
models where the conditional and unconditional probabilities coincide,
\texttt{predict} simulates multiple random networks via the
\texttt{simulate} method in order to estimate the tie probabilities.

In the example below, we use the small \texttt{g4} network with 4 nodes
and 5 directed ties. If we fit a simple ERGM with only the edges
statistic, the maximum likelihood estimate is the logit of 5/12 since
there are \(4\times3=12\) possible ties. If we use the \texttt{predict}
method on this fitted \texttt{ergm} object, theoretically the
conditional and unconditional probabilities are the same because this is
a dyadic independence model. Nonetheless, \texttt{conditional\ =\ FALSE}
forces \texttt{predict} to estimate the matrix of tie probabilities via
simulation of \texttt{nsim} networks.

\begin{Shaded}
\begin{Highlighting}[]
\KeywordTok{data}\NormalTok{(g4)}
\KeywordTok{set.seed}\NormalTok{(}\DecValTok{123}\NormalTok{)}
\NormalTok{SimpleERGM \textless{}{-}}\StringTok{ }\KeywordTok{ergm}\NormalTok{(g4 }\OperatorTok{\textasciitilde{}}\StringTok{ }\NormalTok{edges)}
\KeywordTok{predict}\NormalTok{(SimpleERGM, }\DataTypeTok{conditional =} \OtherTok{TRUE}\NormalTok{, }\DataTypeTok{output =} \StringTok{"matrix"}\NormalTok{)}
\end{Highlighting}
\end{Shaded}

\begin{verbatim}
##           V1        V2        V3        V4
## V1 0.0000000 0.4166667 0.4166667 0.4166667
## V2 0.4166667 0.0000000 0.4166667 0.4166667
## V3 0.4166667 0.4166667 0.0000000 0.4166667
## V4 0.4166667 0.4166667 0.4166667 0.0000000
\end{verbatim}

\begin{Shaded}
\begin{Highlighting}[]
\KeywordTok{predict}\NormalTok{(SimpleERGM, }\DataTypeTok{conditional =} \OtherTok{FALSE}\NormalTok{, }\DataTypeTok{output =} \StringTok{"matrix"}\NormalTok{, }\DataTypeTok{nsim =} \DecValTok{1000}\NormalTok{)}
\end{Highlighting}
\end{Shaded}

\begin{verbatim}
##       V1    V2    V3    V4
## V1 0.000 0.407 0.398 0.434
## V2 0.387 0.000 0.404 0.412
## V3 0.427 0.441 0.000 0.406
## V4 0.416 0.402 0.433 0.000
\end{verbatim}

\hypertarget{flattened-control-arguments-via-a-single-list}{%
\subsection{Flattened control arguments via a single
list}\label{flattened-control-arguments-via-a-single-list}}

Many of the core functions of \pkg{ergm} and related packages have
\texttt{control\ =} arguments that control various aspects of their
working. Within just \pkg{ergm}, for instance, the \texttt{ergm},
\texttt{simulate}, and \texttt{san} functions all require various
control parameters; and packages such as \pkg{ergm.ego} include
additional core functions such as \texttt{ergm.ego}. Moreover, it is not
unusual that, say, a call to \texttt{ergm} will invoke \texttt{simulate}
and possibly even \texttt{SAN} implicitly. This means that a single
\texttt{ergm} (or \texttt{ergm.ego}) call could have multiple lists of
control parameters, sometimes passed as nested lists. \pkg{ergm} 4
implements a method that flattens these nested lists, allowing users to
enter all control parameters in a single list; furthermore, this method
allows for the usual tab-completion of available arguments when using
most R environments.

The key to entering control arguments for all of the various functions
requiring them is the single function \texttt{snctrl()}, which is
shorthand for ``StatNet ConTRoL''. The \texttt{snctrl()} function is
used as the single value of the \texttt{control} argument in a function
such as \texttt{ergm}. For instance, if we wish to force
Monte-Carlo-based estimation in a simple ERGM that could be estimated
exactly---because it is a dyadic independence model in which the
pseudo-likelihood is the same as the likelihood---we might type

\begin{Shaded}
\begin{Highlighting}[]
\KeywordTok{coef}\NormalTok{(}\KeywordTok{ergm}\NormalTok{(g4 }\OperatorTok{\textasciitilde{}}\StringTok{ }\NormalTok{edges, }\DataTypeTok{control =} \KeywordTok{snctrl}\NormalTok{(}\DataTypeTok{force.main =} \OtherTok{TRUE}\NormalTok{, }\DataTypeTok{seed =} \DecValTok{321}\NormalTok{)))}
\end{Highlighting}
\end{Shaded}

\begin{verbatim}
##     edges 
## -0.349256
\end{verbatim}

If the code above is entered in RStudio, then pressing the tab key after
typing ``\texttt{...control\ =\ snctrl(}'' will reveal the various
possible control parameters, including \texttt{force.main}. Additional
illustrations of this method of entering control parameters are in
Krivitsky et al. (2022).

\pkg{ergm} 4 is backwards-compatible with the previous method of passing
control parameters via \texttt{control.ergm}, \texttt{control.simulate},
\texttt{control.san}, and others.

\hypertarget{improved-help-for-model-terms-constraints-and-reference-measures}{%
\subsection{Improved help for model terms, constraints, and reference
measures}\label{improved-help-for-model-terms-constraints-and-reference-measures}}

As alluded to at several points earlier in this article, online help for
model terms, which include term operators, may be obtained by typing
either \texttt{help("{[}name{]}-ergmTerm")} or the shorthand version
\texttt{ergmTerm?{[}name{]}}, where \texttt{{[}name{]}} is the name of
the term or operator. A full list of terms is available via
\texttt{?ergmTerm}, indexed by type and keywords. This list is updated
dynamically as extension packages are loaded and unloaded. Similarly,
help on sample space constraints or reference measures may be obtained
by typing \texttt{?ergmConstraint} or \texttt{?ergmReference},
respectively. Available keywords and their meanings can be obtained by
typing \texttt{?ergmKeywords}. When using RStudio, it is possible to
press the tab key after starting a line with \texttt{?ergm} to view the
wide range of possible help options beginning with the letters
\texttt{ergm}.

\hypertarget{setting-package-options}{%
\subsection{Setting package options}\label{setting-package-options}}

\label{sec:PackageOptions}

\pkg{ergm} 4 has a number of options that affect ERGM estimation as well
as the behavior of some terms, as detailed at the time of this writing
in \secref{sec:global-options} and \ref{sec:term-options}, respectively.
A current list of available options may be obtained via
\texttt{help("ergm-options")} or the shorthand \texttt{options?ergm}.

\hypertarget{sec:global-options}{%
\subsubsection{Global Options}\label{sec:global-options}}

A number of \pkg{ergm} behaviors can be set globally using the familiar
\texttt{options()} command. For example, whether \texttt{ergm()} and
similar functions evaluate the likelihood of the fitted model---a very
computationally intensive process, particularly for valued networks---by
default is controlled by option \texttt{ergm.eval.loglik}, which itself
defaults to \texttt{TRUE}. Running

\begin{Shaded}
\begin{Highlighting}[]
\KeywordTok{options}\NormalTok{(}\DataTypeTok{ergm.eval.loglik =} \OtherTok{FALSE}\NormalTok{)}
\end{Highlighting}
\end{Shaded}

instructs \texttt{ergm()} to skip likelihood calculation unless
overridden in the call via \texttt{ergm(...,\ eval.loglik=TRUE)}.

Other global options currently implemented are

\begin{description}
\item[\texttt{ergm.loglik.warn\_dyads}]
Whether log-likelihood evaluation should issue a warning when the
effective number of dyads that can vary in the sample space is poorly
defined, such as if degree the sequence is constrained.
\item[\texttt{ergm.cluster.retries}]
\pkg{ergm}'s parallel routines implement rudimentary fault-tolerance.
This option controls the number of retries for a cluster call before
giving up.
\item[\texttt{ergm.term}]
This allows the default term options list, described in
\secref{sec:term-options}, to be set globally.
\end{description}

\hypertarget{sec:term-options}{%
\subsubsection{Term options}\label{sec:term-options}}

\pkg{ergm} 4 implements an interface for setting certain options for
ERGM term behavior. The global setting is controlled via
\texttt{options(ergm.term=list(...))} where \texttt{...} are key-value
pairs specifying the options. Individual options can be overwritten on
an ad hoc basis within a function call. For functions that have a
\texttt{control=} argument, such as \texttt{ergm()} and
\texttt{simulate()}, this is done via a \texttt{term.options=} control
parameter, and for those that do not, such as \texttt{summary()}, it is
done by passing the options directly or by passing a
\texttt{term.options=} argument with the list.

Options used as of this writing include:

\begin{description}
\item[\texttt{version}]
A string that can be interpreted as an \proglang{R} package version. If
set, the term will attempt to emulate its behavior as it was that
version of \pkg{ergm}. Not all past version behaviors are available.
\item[\texttt{gw.cutoff}]
In geometrically weighted terms (\texttt{gwesp}, \texttt{gwdegree},
etc.) the highest number of shared partners, degrees, etc. for which to
compute the statistic. This usually defaults to 30.
\item[\texttt{cache.sp}]
Whether the \texttt{gwesp}, \texttt{dgwesp}, and similar terms need
should use a cache for the dyadwise number of shared partners. This
usually improves performance significantly and therefore defaults to
\texttt{TRUE}, but it can be disabled.
\item[\texttt{interact.dependent}]
How to handle attempts to use interaction terms \texttt{:} and
\texttt{*} with dyad-dependent terms. Possible values are
\texttt{"error"} (the default), \texttt{"message"}, \texttt{"warning"},
and \texttt{"silent"}. Each of the last three will allow such terms,
defined as described in \secref{sec:InteractionEffects} via their change
statistics.
\end{description}

\hypertarget{discussion}{%
\section{Discussion}\label{discussion}}

Since version 2.1 of the \pkg{ergm} package was released concurrently
with Hunter, Handcock, et al. (2008), the package has undergone
substantial changes. This paper describes the changes that are most
likely to be of general interest, including---but not limited to---those
that are new with the release of major version 4 (Handcock et al.,
2021). Development of \pkg{ergm} and the growing list of related
packages, many of which are described in \secref{sec:Extensions} of this
article, is ongoing. Thus, while this article describes many new
features, it represents a snapshot of the evolving code comprising the
\statnet{} suite of packages for \proglang{R} (\proglang{R} Core Team,
2021).

\hypertarget{acknowledgments}{%
\section*{Acknowledgments}\label{acknowledgments}}
\addcontentsline{toc}{section}{Acknowledgments}

Many individuals have contributed code for version 4 of \pkg{ergm},
particularly Mark Handcock, who wrote most of the code upon which
missing data inference and diagnostics are based, and Michał Bojanowski,
who produced the \texttt{predict} method, among many other contributions
by both of them. Carter Butts is the main developer of the \pkg{network}
package, upon which \pkg{ergm} depends; in addition, he provided
numerous suggestions for computational improvements and new terms, and
provided numerous helpful comments about this manuscript. Skye
Bender-deMoll wrote a vignette that automatically cross-references
\texttt{ergm} model terms, Joyce Cheng wrote the dynamic documentation
system and miscellaneous enhancements, and Christian Schmid contributed
code improving MPLE standard error estimation. Other important
contributors are Steven Goodreau, Ayn Leslie-Cook, Li Wang, and Kirk Li.
We are grateful to all these individuals as well as the many users of
\pkg{ergm} who have aided the package's development through the many
questions and suggestions they have posed over the years.

\hypertarget{references}{%
\section*{References}\label{references}}
\addcontentsline{toc}{section}{References}

\hypertarget{refs}{}
\begin{cslreferences}
\leavevmode\hypertarget{ref-network.dynamic}{}%
Bender-deMoll, S. (2016). \emph{Temporal network tools in statnet:
networkDynamic, ndtv And tsna}. Statnet Development Team.
\url{http://statnet.org/Workshops/ndtv_workshop.html}

\leavevmode\hypertarget{ref-Bu08s}{}%
Butts, C. T. (2008). Social network analysis with sna. \emph{Journal of
Statistical Software}, \emph{24}(6), 1--51.
\url{https://doi.org/10.18637/jss.v024.i06}

\leavevmode\hypertarget{ref-caimo2020}{}%
Caimo, A., \& Gollini, I. (2020). A multilayer exponential random graph
modelling approach for weighted networks. \emph{Computational Statistics
and Data Analysis}, \emph{142}, 106825.
\url{https://doi.org/10.1016/j.csda.2019.106825}

\leavevmode\hypertarget{ref-coleman1964}{}%
Coleman, J. S. (1964). \emph{Introduction to mathematical sociology}.
The Free Press of Glencoe.

\leavevmode\hypertarget{ref-R2021}{}%
\proglang{R} Core Team. (2021). \emph{R: A language and environment for
statistical computing}. R Foundation for Statistical Computing.
\url{http://www.R-project.org/}

\leavevmode\hypertarget{ref-HaGi10m}{}%
Handcock, M. S., \& Gile, K. J. (2010). Modeling social networks from
sampled data. \emph{Annals of Applied Statistics}, \emph{4}(1), 5--25.
\url{https://doi.org/10.1214/08-AOAS221}

\leavevmode\hypertarget{ref-ergm4}{}%
Handcock, M. S., Hunter, D. R., Butts, C. T., Goodreau, S. M.,
Krivitsky, P. N., \& Morris, M. (2021). \emph{ergm: Fit, simulate and
diagnose exponential-family models for networks}. The Statnet Project
(\url{https://statnet.org}).
\url{https://CRAN.R-project.org/package=ergm}

\leavevmode\hypertarget{ref-HeWi20p}{}%
Henry, L., \& Wickham, H. (2020). \emph{purrr: Functional programming
tools}. \url{https://CRAN.R-project.org/package=purrr}

\leavevmode\hypertarget{ref-holland1981}{}%
Holland, P. W., \& Leinhardt, S. (1981). An exponential family of
probability distributions for directed graphs. \emph{Journal of the
American Statistical Association}, \emph{76}(373), 33--50.

\leavevmode\hypertarget{ref-HuHu12i}{}%
Hummel, R. M., Hunter, D. R., \& Handcock, M. S. (2012). Improving
simulation-based algorithms for fitting ERGMs. \emph{Journal of
Computational and Graphical Statistics}, \emph{21}(4), 920--939.
\url{https://doi.org/10.1080/10618600.2012.679224}

\leavevmode\hypertarget{ref-ergm.userterms}{}%
Hunter, D. R., \& Goodreau, S. M. (2019). \emph{Extending ergm
functionality within statnet: Building custom user terms}. Statnet
Development Team.
\url{http://statnet.org/Workshops/ergm.userterms_tutorial.pdf}

\leavevmode\hypertarget{ref-HuGo08g}{}%
Hunter, D. R., Goodreau, S. M., \& Handcock, M. S. (2008). Goodness of
fit for social network models. \emph{Journal of the American Statistical
Association}, \emph{103}(481), 248--258.
\url{https://doi.org/10.1198/016214507000000446}

\leavevmode\hypertarget{ref-HuGo13e}{}%
Hunter, D. R., Goodreau, S. M., \& Handcock, M. S. (2013).
ergm.userterms: A template package for extending statnet. \emph{Journal
of Statistical Software}, \emph{52}(2), 1--25.
\url{https://doi.org/10.18637/jss.v052.i02}

\leavevmode\hypertarget{ref-HuHa08e}{}%
Hunter, D. R., Handcock, M. S., Butts, C. T., Goodreau, S. M., \&
Morris, M. (2008). ergm: A package to fit, simulate and diagnose
exponential-family models for networks. \emph{Journal of Statistical
Software}, \emph{24}(3), 1--29.
\url{https://doi.org/10.18637/jss.v024.i03}

\leavevmode\hypertarget{ref-epimodel}{}%
Jenness, S. M., Goodreau, S. M., \& Morris, M. (2018). EpiModel: An R
package for mathematical modeling of infectious disease over networks.
\emph{Journal of Statistical Software}, \emph{84}(8), 1--47.
\url{https://doi.org/10.18637/jss.v084.i08}

\leavevmode\hypertarget{ref-KaKr17s}{}%
Karwa, V., Krivitsky, P. N., \& Slavković, A. B. (2017). Sharing social
network data: Differentially private estimation of exponential-family
random graph models. \emph{Journal of the Royal Statistical Society,
Series C}, \emph{66}(3), 481--500.
\url{https://doi.org/10.1111/rssc.12185}

\leavevmode\hypertarget{ref-Kr12e}{}%
Krivitsky, P. N. (2012). Exponential-family random graph models for
valued networks. \emph{Electronic Journal of Statistics}, \emph{6},
1100--1128. \url{https://doi.org/10.1214/12-EJS696}

\leavevmode\hypertarget{ref-Kr20r}{}%
Krivitsky, P. N. (2020). \emph{rle: Common functions for run-length
encoded vectors}. \url{https://CRAN.R-project.org/package=rle}

\leavevmode\hypertarget{ref-KrBu17e}{}%
Krivitsky, P. N., \& Butts, C. T. (2017). Exponential-family random
graph models for rank-order relational data. \emph{Sociological
Methodology}, \emph{47}(1), 68--112.
\url{https://doi.org/10.1177/0081175017692623}

\leavevmode\hypertarget{ref-valued}{}%
Krivitsky, P. N., \& Butts, C. T. (2019). \emph{Modeling valued networks
with statnet}. Statnet Development Team.
\url{http://statnet.org/Workshops/valued.html}

\leavevmode\hypertarget{ref-KrHa08}{}%
Krivitsky, P. N., \& Handcock, M. S. (2008). Fitting position latent
cluster models for social networks with latentnet. \emph{Journal of
Statistical Software}, \emph{24}(5), 1--23.
\url{https://doi.org/10.18637/jss.v024.i05}

\leavevmode\hypertarget{ref-KrHa14}{}%
Krivitsky, P. N., \& Handcock, M. S. (2014). A separable model for
dynamic networks. \emph{Journal of the Royal Statistical Society, Series
B}, \emph{76}(1), 29--46. \url{https://doi.org/10.1111/rssb.12014}

\leavevmode\hypertarget{ref-krivitsky2009}{}%
Krivitsky, P. N., Handcock, M. S., Raftery, A. E., \& Hoff, P. D.
(2009). Representing degree distributions, clustering, and homophily in
social networks with latent cluster random effects models. \emph{Social
Networks}, \emph{31}(3), 204--213.
\url{https://doi.org/10.1016/j.socnet.2009.04.001}

\leavevmode\hypertarget{ref-ergm4b}{}%
Krivitsky, P. N., Hunter, D. R., Morris, M., \& Klumb, C. (2022).
\emph{ergm 4: Computational improvements}.
\url{https://arxiv.org/abs/TBD}

\leavevmode\hypertarget{ref-KrKo20e}{}%
Krivitsky, P. N., Koehly, L. M., \& Marcum, C. S. (2020).
Exponential-family random graph models for multi-layer networks.
\emph{Psychometrika}, \emph{85}, 630--659.
\url{https://doi.org/10.1007/s11336-020-09720-7}

\leavevmode\hypertarget{ref-KrMo17i}{}%
Krivitsky, P. N., \& Morris, M. (2017). Inference for social network
models from egocentrically-sampled data, with application to
understanding persistent racial disparities in HIV prevalence in the US.
\emph{Annals of Applied Statistics}, \emph{11}(1), 427--455.
\url{https://doi.org/10.1214/16-AOAS1010}

\leavevmode\hypertarget{ref-MoHa08spec}{}%
Morris, M., Handcock, M. S., \& Hunter, D. R. (2008). Specification of
exponential-family random graph models: Terms and computational aspects.
\emph{Journal of Statistical Software}, \emph{24}(4), 1--24.
\url{https://doi.org/10.18637/jss.v024.i04}

\leavevmode\hypertarget{ref-tergm}{}%
Morris, M., \& Krivitsky, P. N. (2015). \emph{Temporal exponential
random graph models (TERGMs) for dynamic network modeling in statnet}.
Statnet Development Team.
\url{http://statnet.org/Workshops/tergm_tutorial.html}

\leavevmode\hypertarget{ref-ergm.ego}{}%
Morris, M., \& Krivitsky, P. N. (2019). \emph{Introduction to egocentric
network data analysis with ERGMs using statnet}. Statnet Development
Team. \url{http://statnet.org/Workshops/ergm.ego_tutorial.html}

\leavevmode\hypertarget{ref-RoPa99l}{}%
Robins, G., Pattison, P., \& Wasserman, S. (1999). Logit models and
logistic regressions for social networks: III. Valued relations.
\emph{Psychometrika}, \emph{64}(3), 371--394.

\leavevmode\hypertarget{ref-Sa68n}{}%
Sampson, S. F. (1968). \emph{A novitiate in a period of change: An
experimental and case study of social relationships} {[}Ph.D.~thesis
(University Micofilm, No 69-5775){]}. Department of Sociology, Cornell
University.

\leavevmode\hypertarget{ref-schmid2020}{}%
Schmid, C. S., \& Hunter, D. R. (2020). \emph{Improving ERGM starting
values using simulated annealing}.
\url{https://arxiv.org/abs/2009.01202}

\leavevmode\hypertarget{ref-schweinberger2020}{}%
Schweinberger, M., Krivitsky, P. N., Butts, C. T., \& Stewart, J. R.
(2020). Exponential-family models of random graphs: Inference in finite,
super and infinite population scenarios. \emph{Statistical Science},
\emph{35}(4), 627--662. \url{https://doi.org/10.1214/19-STS743}

\leavevmode\hypertarget{ref-SlKo16m}{}%
Slaughter, A. J., \& Koehly, L. M. (2016). Multilevel models for social
networks: Hierarchical Bayesian approaches to exponential random graph
modeling. \emph{Social Networks}, \emph{44}, 334--345.
\url{https://doi.org/10.1016/j.socnet.2015.11.002}

\leavevmode\hypertarget{ref-vegayon2021}{}%
Vega Yon, G. G., Slaughter, A., \& de la Haye, K. (2021). Exponential
random graph models for little networks. \emph{Social Networks},
\emph{64}, 225--238. \url{https://doi.org/10.1016/j.socnet.2020.07.005}

\leavevmode\hypertarget{ref-Wa12e}{}%
Wang, P. (2012). Exponential random graph model extensions: Models for
multiple networks and bipartite networks. In D. Lusher, J. Koskinen, \&
G. Robins (Eds.), \emph{Exponential random graph models for social
networks: Theory, methods, and applications} (pp. 115--129). Cambridge
University Press. \url{https://doi.org/10.1017/CBO9780511894701.012}
\end{cslreferences}

\end{document}